\documentclass[twocolumn,superscriptaddress,prb, amsmath,amssymb, aps]{revtex4-2}
\usepackage{xcolor}
\usepackage{graphicx}
\usepackage{dcolumn}
\usepackage{bm}
\usepackage{svg}
\usepackage{float}
\newcommand{\bs}[1]{{\boldsymbol{#1}}}
\newcommand{\bk}{\bs{k}}
\newcommand{\br}{\bs{r}}
\newcommand{\bq}{\bs{q}}

\usepackage{cancel}
\usepackage{soul}
\usepackage[caption=false]{subfig}
\usepackage{hyperref}

\begin{document}

\title{Entanglement entropy across the dynamical phase transition \\ in the quantum $\mathcal{O}(N)$ model}

\author{Frederick Del Pozo}
\email{frederick.delpozo@lkb.upmc.fr}
\affiliation{Laboratoire Kastler Brossel, Sorbonne Universit\'{e}, CNRS, ENS-PSL Research University, 
Coll\`{e}ge de France; 4 Place Jussieu, 75005 Paris, France}
\author{Tangi Morvan}
\email{tangi.morvan@universite-paris-saclay.fr}
\affiliation{Laboratoire de Physique Théorique et Modèles Statistiques, Batiment 530, Rue André Rivière, 91405 Orsay,
France}
\affiliation{Ecole Normale Supérieure de Lyon, 15 parvis René Descartes, 69342 Lyon, France}

\author{Ir\'en\'ee Fr\'erot}
\email{irenee.frerot@lkb.upmc.fr}
\affiliation{Laboratoire Kastler Brossel, Sorbonne Universit\'{e}, CNRS, ENS-PSL Research University, 
Coll\`{e}ge de France; 4 Place Jussieu, 75005 Paris, France}

\author{Nicolas Cherroret}
\email{nicolas.cherroret@lkb.upmc.fr}  
\affiliation{Laboratoire Kastler Brossel, Sorbonne Universit\'{e}, CNRS, ENS-PSL Research University, 
Coll\`{e}ge de France; 4 Place Jussieu, 75005 Paris, France}

\begin{abstract}
We demonstrate that the dynamical phase transition of the quantum $\mathcal{O}(N)$ model at large $N$ leaves universal fingerprints in the infrared structure of the entanglement spectrum. 
While the leading contribution to the entanglement entropy at long time follows the conventional volume law associated with ballistic entanglement spreading, its subleading behavior sharply distinguishes the different dynamical regimes.
Specifically, quenches at and below the critical point generate gapless low-energy entanglement modes together with logarithmic corrections to the long-time entanglement entropy, whose scaling is governed by the scaling exponent of the transition. 
Using an infinite-slab bipartition geometry and exact numerical correlation functions in the large-$N$ limit, we characterize these scaling laws across the dynamical phase diagram and relate them to the emergence of long-range correlations during the post-quench dynamics. We further show that the entanglement eigenmodes themselves reveal characteristic signatures of the dynamical phase transition through their spatio-temporal structure and degeneracy properties.
\end{abstract}

\keywords{Suggested keywords}
                         
\maketitle

\clearpage
\section{Introduction}\label{sec:intro}

The far-from-equilibrium dynamics of quantum many-body systems across a phase transition exhibits a rich phenomenology that goes beyond the conventional framework of equilibrium critical phenomena. While this dynamics generally reflects universal features tied to the symmetries of the underlying phase transition, it also displays genuinely non-equilibrium scaling behavior governed by additional ingredients such as conservation laws or the dynamics of topological defects \cite{Hohenberg1997}. In this context, over the past decade a growing number of experimental platforms have explored the scaling laws emerging after quantum quenches across phase transitions. Notable examples include quenches of cold-atomic gases across Ising-like phase transitions \cite{Huh2024, Manovitz2025}, as well as cooling and heating quenches across condensation \cite{Erne2018, Glidden2021} and Kosterlitz–Thouless transitions \cite{Abuzarli2022, Sunami2023, Gazo2025}.
This experimental progress has been accompanied by intense theoretical efforts devoted to the non-equilibrium dynamics of quantum gases in a wide range of models \cite{Berges2008, Nowak2012, Chantesana2019, Mikheev2019,Regemortel2018, Schmied2019, Comaron2019, Duval2023, Gliott2024, Duval2025, Alilou2025, paper:quench_prot}.

Quantum quenches of many-body systems across an equilibrium phase transition may give rise to the concept of \emph{dynamical} phase transition (DPT): depending on whether the post-quench energy lies in the ordered or disordered phase, or near the critical point of the associated equilibrium problem, correlations may spontaneously develop distinct dynamical scaling laws. In general, the scaling behavior following quenches into an ordered phase is associated with coarsening dynamics, where ordered regions grow across increasing length scales through the recombination and annihilation of topological defects {\color{black}\cite{Bray1993, Berthier2001, Cugliandolo2015, Gliott2025, Balducci2026}}. Although a rigorous theoretical description of DPTs remains challenging in most cases, the quantum $\mathcal{O}(N)$ model \cite{Moshe2003} has recently attracted considerable attention in this context. In the large-$N$ limit, it was shown to exhibit a DPT, with correlation functions displaying coarsening dynamics for quenches below or at the critical point in spatial dimensions $d \geq 3$ \cite{Chandran2013, Sciolla2013, Maraga2015, Smacchia2015, Diehl2017, Halimeh2021, Cherroret2024}.
In addition to exhibiting a rich dynamical phase diagram, the DPT in the $\mathcal{O}(N)$ model at large $N$ is governed by an effective quadratic theory, making it amenable to analytical treatment. 
A direct consequence is that the underlying dynamics is integrable, a property recently formalized within the framework of generalized Gibbs ensembles \cite{Giachetti2025}. Although this integrability may prevent the model from capturing the true long-time behavior of DPTs in generic non-integrable systems, it nevertheless provides a valuable paradigmatic setting for studying DPTs in prethermal regimes, where the dynamics can be approximately described by an effective integrable theory.

While the dynamical properties of correlations in the quantum $\mathcal{O}(N)$ model have been extensively investigated over the past decade, the corresponding dynamics of \emph{quantum entanglement} in the vicinity of the DPT has received far less attention. In equilibrium systems near a quantum phase transition, it is well known that the entanglement entropy (EE) $S$ carries clear signatures of criticality. For instance, in one-dimensional quantum critical systems, the area law is violated in favor of a logarithmic scaling with the subsystem size $L$, namely $S \propto \ln L$ \cite{Laflorencie2016, Turner2011}. In higher dimensions, the EE generally follows an area law, $S = a L^{d-1}$, but critical behavior still emerges through a cusp singularity in the prefactor $a$ as the system crosses a quantum phase transition \cite{Metlitski2009, Casini2012, Kallin2013, Helmes2014, Frerot2016}.
In \emph{out-of-equilibrium} many-body systems, quantum quenches typically lead to a rapid production of entanglement entropy, a phenomenon that has attracted considerable attention \cite{Peschel2008, Calabrese2007, Poilblanc2011, Frerot2018}. In generic situations, the ensuing dynamics is characterized by a light-cone structure and the emergence of a volume law at long times. To what extent this behavior is modified by the presence of a DPT, however, remains largely unexplored. Addressing this question is the main objective of the present work, where we use the DPT in the quantum $\mathcal{O}(N)$ model as a representative example.
Specifically, while we do not observe any pronounced signature of the DPT in the dynamics of EE at leading order, we find that it develops critical subleading corrections to the volume law at long times. For quenches at or below the critical point, these corrections take the form of time-dependent logarithmic contributions to the EE, whose behavior further depends on the scaling exponent $\alpha$ governing correlations around the DPT. To uncover these corrections, we investigate the dynamics of the low-lying modes of the entanglement spectrum (ES) \cite{Laflorencie2016}, which encodes more detailed information than the EE itself, as is known for example in the context of topological phase transitions \cite{paper:OG_ESpec, paper:herviou2016, paper:delpozo1, paper:delpozo2} and in systems with a spontaneously broken continuous symmetry \cite{Melitski2015,Frerot2015,Frerot2017}.
The entanglement spectrum of the DPT in the $\mathcal{O}(N)$ model was previously studied in the seminal work of Ref.~\cite{Lemonik2016}, albeit using approximate expressions for the correlation functions that are valid only for very deep quenches {\color{black}and focusing exclusively on quenches to the critical point. Moreover, the dynamical scaling laws for the EE derived in Ref.~\cite{Lemonik2016} were obtained for relatively small subsystems, a regime in which the logarithmic corrections discussed above are not visible. To overcome these limitations and elucidate the sensitivity of the ES to the underlying DPT, we adopt a different approach in the present work. Specifically, we study the EE across the three-dimensional DPT of the $\mathcal{O}(N)$ model for \emph{both} quenches below and to the critical point, using an \emph{infinite-slab} bipartition geometry.} This setup allows us to probe the thermodynamic limit of asymptotically long times for both the ES and the EE, while performing numerical simulations based on the exact correlation functions of the $\mathcal{O}(N)$ model. By analyzing quenches above, at, and below the critical point, we identify the dynamical scaling laws governing the entanglement spectrum in each regime and, in turn, the critical subleading contributions to the EE. 

The paper is organized as follows. In Sec.~\ref{sec:theory_field_theory}, we briefly review the main properties of the DPT in the $\mathcal{O}(N)$ model in the large-$N$ limit, following a quench starting from the disordered phase.
Section~\ref{sec:theory_ent_spec} presents the numerical framework used to compute the entanglement spectrum and entanglement entropy in the infinite-slab geometry considered throughout this work. 
We then analyze the low-energy structure of the entanglement spectrum across the three quench regimes in Sec.~\ref{sec:results}, and derive the associated logarithmic corrections to the entanglement entropy. 
In Sec.~\ref{sec:ball}, we investigate the dynamics and spatial structure of the lowest entanglement modes, highlighting further signatures of the dynamical phase transition. Finally, Sec.~\ref{sec:conclusion} summarizes our results and discusses possible future directions. {\color{black}Details of the numerical and fitting strategies, as well as additional scaling analyses, are presented in three appendices.}
\section{Quench dynamics in the $\mathcal{O}(N)$ model}
\label{sec:theory_field_theory}

The focus of this work is the Hamiltonian of $\mathcal{O}(N)$ bosons with quartic interactions in three spatial dimensions ($d = 3$): 
\begin{equation}
\label{eq:Ham_def_model}
    H = \frac{1}{2}\int \mathrm{d}^{3}\br \!\left[{\bm{\Pi}}^{{\color{black} \mathrm{T}}}{\bm{\Pi}} \!+\! {\bm{\phi}}^{{\color{black} \mathrm{T}}}\left(r- \nabla^{2}\! +\! \frac{{\color{black} 2}u}{4!N}|{\bm{\phi}}|^{2}  \right){\bm{\phi}}     \right].
\end{equation}
Here ${\bm{\phi}}$ 
and $\bm{\Pi}$ {\color{black} are real, $N$-component 
} canonically conjugate fields, satisfying $[\phi_i(\br),\Pi_j(\br')]=i\delta_{ij}\delta(\br-\br')$. The parameters $r$ and $u$ denote the ``mass'' and interaction strength, respectively. 
In the following, we study the dynamics of the $\mathcal{O}(N)$ model in the limit $N\to\infty$, where it becomes solvable via a self-consistent Gaussian theory \cite{Chandran2013, Smacchia2015, Maraga2015}. In this limit, the state of the system at any time $t$ reduces to a Gaussian product state over the component indices, implying the large-$N$ factorization 
$\bm{\phi}^{{\color{black} \mathrm{T}}}\bm{\phi}\,\phi_i=\langle {\bm{\phi}}^{{\color{black} \mathrm{T}}}{\bm{\phi}}\rangle \phi_i$. For dynamics initialized in the disordered phase of the equilibrium model, as assumed throughout this work, the mean field vanishes at all times $\langle\phi_i\rangle=0$.
Under these conditions, the Heisenberg equations of motion for each component $\phi_i$ (hereafter denoted simply by $\phi$), reduce to $\mathrm{d}_t \phi=\Pi$ and $\mathrm{d}_t\Pi=-\delta H_\text{eff}/\delta\phi$, with the effective time-dependent Hamiltonian
\begin{equation}\label{eq:effective_Ham}
H_{\mathrm{eff}}(t)= \frac{1}{2}\int\mathrm{d}^{3}\br \left[|\Pi|^{2} + \phi\left(r_\text{eff}(t) - \nabla^{2}\right)\phi \right].
\end{equation}
This Hamiltonian is now effectively quadratic, and involves a time-dependent effective mass 
{\color{black} $r_\text{eff}(t)=r+
\frac{u}{6N}
\left\langle
{\bm{\phi}}^{\mathrm{T}}{\bm{\phi}}
\right\rangle.$
Using the $\mathcal{O}(N)$ symmetry of the state, we have $\left\langle
{\bm{\phi}}^{\mathrm{T}}{\bm{\phi}}
\right\rangle
=\sum_{i=1}^{N}\langle\phi_i^2\rangle=
N\langle\phi^2\rangle,$ with $\phi$ any arbitrary component, such that the effective mass can equivalently be written as $r_\text{eff}(t)=r+
\frac{u}{6}\langle\phi^2\rangle.$
In what follows, we study the post-quench dynamics of such a single component, with extensive quantities such as the energy and entanglement entropy of the full system scaling as $N$ times the corresponding quantity for a single component.}

The equations of motion are conveniently simplified by decomposing the time-dependent Fourier component $\phi_\bk(t)$ in terms of creation and annihilation operators $\smash{a_\bk^\dagger}$ and $a_\bk$ as $\smash{\phi_\bk(t) = f_\bk(t) a_\bk+ {f}^*_{\bk}(t)a^{\dagger}_{-\bk}}$, where the mode functions satisfy
$f_{-\bk}(t)=f_\bk(t)$ and the Wronskian condition $2\mathfrak{Im}[f_\bk(t)\dot{f}_\bk^*(t)] = 1$, imposed by the bosonic commutation relations. Inserting this into Eq. \eqref{eq:effective_Ham} leads to the equation of motion
\begin{equation}
\label{eq:mode_func_eom}
  \left[\partial_t^2+ k^2 + r_\text{eff}(t)\right] f_\bk(t) = 0,
\end{equation}
with the effective mass given by
\begin{equation}
\label{eq:effective_r}
    r_\text{eff}(t) = r + \frac{u}{6} \int \frac{\mathrm{d}^3\bk
    }{(2\pi)^3} \mathcal{R}_{\Lambda}(k)|f_\bk(t)|^2.
\end{equation}
Here, we introduced a regulator function $\mathcal{R}_\Lambda(k)$  that suppresses high-momentum contributions $k>\Lambda$. While such large momenta have no impact on the dynamical and critical exponents of the DPT underlying Eq. \eqref{eq:Ham_def_model}, which are governed by the infrared behavior of correlation functions, they do lead to non-universal finite-size effects \cite{Giachetti2025}. In particular, it was shown in \cite{Maraga2015} that sharp regulators make the extraction of universal features of the DPT more difficult due to persistent finite-size effects at long times. For this reason, in all subsequent simulations we employ a smooth Gaussian regulator, $\smash{\mathcal{R}_\Lambda(k) = e^{-k^{2}/(2\Lambda^{2})}}$ with $\Lambda = \pi/2$. 
As an initial condition for Eq.~\eqref{eq:mode_func_eom}, we consider the zero-temperature (and zero-entropy) ground state of the non-interacting Hamiltonian ($u =0$), with initial mass $r_\text{eff}(t=0) = r_i$. 
This implies $\smash{\langle a^{\dagger}_{\bk}a_{\bk}\rangle = 0}$ and initial mode functions satisfying $\smash{f_{\bk}(t=0) = 1/(\sqrt{2\omega_{\bk}^{0}})}$ and $\smash{\dot{f}_{\bk}(0) = -i\sqrt{\omega^{0}_{\bk}/2}}$, with $\omega^{0}_{\bk} = \sqrt{k^{2} + r_i}$.

\begin{figure}[h]
    \centering
    \includegraphics[width=0.9\linewidth]{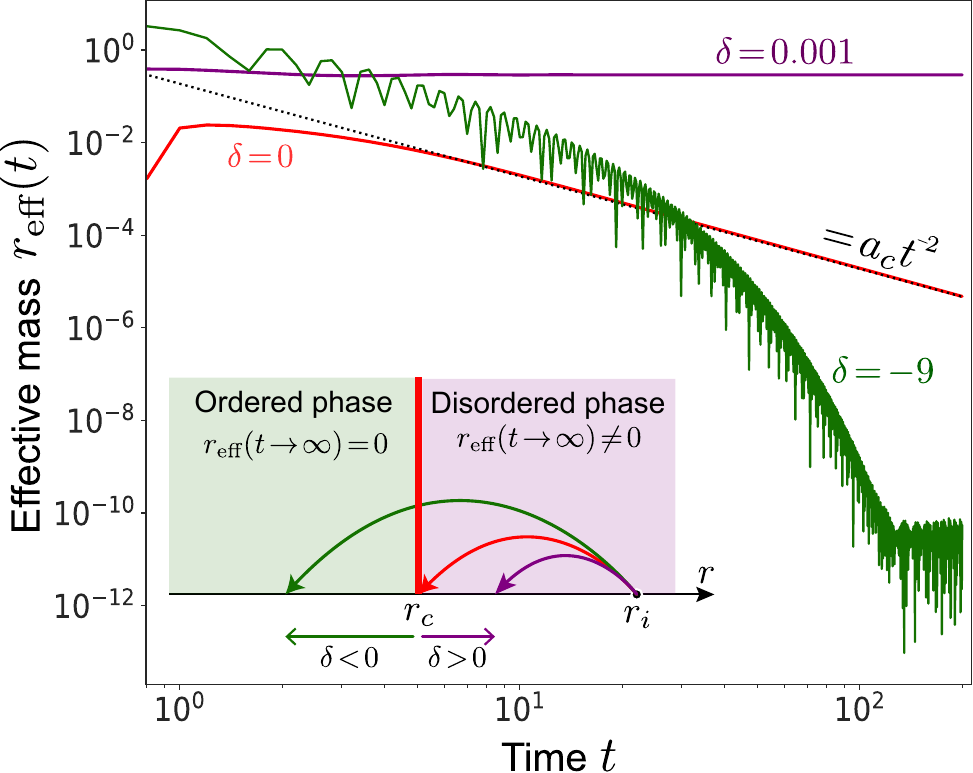}
    \caption{Effective mass $r_\text{eff}(t)$ in dimension $d=3$, obtained by numerically solving Eqs. \eqref{eq:mode_func_eom} and \eqref{eq:effective_r} for $\delta=0.001$ (purple curve), $\delta=0$ (red curve) and $\delta=-9$ (green curve), corresponding to quenches above, at, and below the critical point, respectively. Here we choose $r_{i} = 100$ and $u=10$, yielding a critical value $r_c\simeq-0.43$. The numerical simulations are performed for $k\in \left[0, 10\Lambda\right]$ and $10^{5}$ points. The dashed line indicates the asymptotic behavior $r_\text{eff}(t)= a_{c}t^{-2}$, where $a_{c} = \frac{d}{4}(1 - \frac{d}{4}) =  3/16$.
    The inset shows a schematic illustration of the dynamical phase transition in three dimensions, following a sudden quench of the mass parameter from an initial value $r_{i}$  to a final value  $r$. Quenches with $r>r_c$, where  $r_{c}$ is defined in Eq. (\ref{eq:rc}), lead to a finite asymptotic value of the effective mass $r_\text{eff}(t)$ at  long times. In contrast, for quenches with $r\le r_c$, the effective mass decays to zero at long time.}
    \label{fig:DPT}
\end{figure}

Starting from this initial condition, the coupled equations \eqref{eq:mode_func_eom} and \eqref{eq:effective_r} are known to exhibit a DPT when the mass parameter  is quenched from $r_i$ to $r$ below a critical value $r_c$, as  illustrated in the inset of Fig.~\ref{fig:DPT}: For $r\leq r_c$, the effective mass $r_\text{eff}(t)$ vanishes at long times, so that Eq.~\eqref{eq:mode_func_eom} becomes scale invariant and the system’s correlation functions exhibit self-similar scaling, or \emph{coarsening}. The critical value $r_c$ is negative and is given by \cite{Smacchia2015, Maraga2015}
\begin{equation}
\label{eq:rc}
    r_{c} = -\frac{u}{24}\int \frac{\mathrm{d}^3\bk
    }{(2\pi)^3} \mathcal{R}_{\Lambda}(k) \frac{2k^{2} + r_i}{k^{2}\sqrt{k^{2} + r_i}}.
\end{equation}
The DPT is controlled by the quench depth $\delta \equiv (r-r_c)/|r_{c}|$, whose value governs the late-time behavior of the effective mass $r_\text{eff}(t)$ and, in turn, that of the correlation functions. For quenches above the critical point ($\delta > 0$), $r_\text{eff}(t \to \infty)$ approaches a finite constant which, close to criticality, scales as $|\delta|$ \cite{Diehl2017}. In contrast, at and below the critical quench ($\delta = 0$ and $\delta < 0$, respectively), $r_\text{eff}(t)$ vanishes asymptotically. Previous works \cite{Maraga2015} have shown that, in these latter two cases, $r_\text{eff}(t) = a/t^2$, with $a = (3 - d)(d - 1)/4$ for $\delta < 0$, and $a = a_c = (d/4)(1 - d/4)$ for a critical quench. 
These behaviors are illustrated in Fig.~\ref{fig:DPT}, which displays the time evolution of $r_\text{eff}(t)$ numerically computed from Eqs. (\ref{eq:mode_func_eom}) and (\ref{eq:effective_r}) for the three types of quenches in dimension $d=3$  {\color{black}(see Appendix \ref{app:num_methods} for details about the numerical methods)}.
In particular, we do observe a decay compatible with {\color{black}$r_\text{eff}(t)=3/(16t^{2})$} at the critical quench, while for $r<r_c$ the effective mass vanishes exponentially, in accordance with the prediction that $a=0$.

The different asymptotic behavior of $r_\text{eff}(t)$ across the DPT also leads to qualitatively distinct dynamical scaling laws for two-point correlation functions. In particular, this is the case for the field-field correlation function in momentum space, defined as 
\begin{equation}
\label{eq:phiphicor}
G_{\phi\phi}(\bk,t)\equiv
\langle\phi_\bk(t)\phi^\dagger_\bk(t)\rangle\equiv |f_\bk(t)|^2.
\end{equation}
Following a quench below or at the critical point, it takes the self-similar form
\begin{equation}
\label{eq:self-similar}
G_{\phi\phi}(\bk,t)=
t^{2\alpha+1}f(kt^z).
\end{equation}
{\color{black}Here $z=1$ is the dynamical exponent of the DPT, while the \emph{scaling exponent} \footnote{$\alpha$ is related to the so-called \emph{slip-exponent} $\theta$ in previous works \cite{Maraga2015, Diehl2017}.}} is     $\alpha=(d-2)/2$ for a quench below $r_c$, and $\alpha=\alpha_c=(d-2)/4$ at criticality ($r=r_c$). The explicit form of the scaling function $f$ was derived in \cite{Maraga2015} for the particular case of very deep quenches, i.e., for an initial state such that $r_i \gg r_c$. At long times, corresponding to $x = kt \gg 1$, the authors of \cite{Maraga2015} found $f(x)\sim 1/x^{2\alpha+1}$, which yields
\begin{equation}
\label{eq:Gphiphi_lowk}
G_{\phi\phi}(\bk,t\gg k^{-1})\sim \frac{1}{k^{2\alpha+1}}.
\end{equation}
For quenches below $r_c$, this gives for instance $G_{\phi\phi}(\bk,t)\sim 1/k^{d-1}$, generally different from a thermal decay $\sim 1/k^2$. We have verified that this characteristic decay actually holds for arbitrary quench depths, for which no analytical prediction is currently available.
Note that the non-thermal decay $G_{\phi\phi}(\bk,t)\sim 1/k^{d-1}$ highlights the integrable nature of the $\mathcal{O}(N)$ model in the large-$N$ limit. As such, the system is expected to asymptotically approach a generalized Gibbs ensemble (GGE) at late times. The exact structure of this GGE was only recently determined in \cite{Giachetti2025}.

\section{Entanglement entropy in the quantum $\mathcal{O}(N)$ model}
\label{sec:theory_ent_spec}

\subsection{Definition and Williamson approach}

The EE $S(A|B)$ between a subregion $A$ and its complement $B$ is defined as \cite{Peschel2009}
\begin{equation}
    S(A|B) = -\mathrm{tr}_{A}\left[ \rho_{A} \ln\left(\rho_{A}\right) \right],
\end{equation}
where 
$\rho_{A} \equiv \mathrm{tr}_{B}\left( \rho\right)$, with $\rho=|\Psi(t)\rangle\langle\Psi(t)|$ the density matrix of the full system.
The reduced density matrix can always be expressed as $\rho_{A} \propto e^{-\mathcal{H}_{E}}$, thereby defining the entanglement Hamiltonian $\mathcal{H}_{E}$.
In practice, the evaluation of $S(A|B)$  thus reduces to computing the spectrum of $\mathcal{H}_{E}$.
This computation is greatly simplified for Gaussian states, as is the case in the $\mathcal{O}(N)$ model at large $N$, for which the entanglement Hamiltonian is quadratic in the fields $\phi$ and $\Pi$, and therefore can be diagonalized as a set of non-interacting bosonic modes via a Bogoliubov transformation \cite{Frerot2015}. The entanglement Hamiltonian takes the form $\smash{{\cal H}_E \equiv\sum_j \omega_j b_j^\dagger b_j}$ with $b_j$ bosonic modes capturing the entanglement structure, and $\omega_j\ge 0$ the (one-body) entanglement spectrum.
To obtain the entanglement spectrum, a convenient strategy is provided by the Williamson approach \cite{Peschel2009, Botero2004, Frerot2015}, which relates the $\omega_j$ to the symplectic spectrum $\{\lambda_j\}$ of the correlation matrix $G \equiv \frac{1}{2}\mathrm{tr}_{A}\left( \{\eta,\eta^{\dagger}\} \rho_{A}\right) $, where $\eta = \left(\phi, \Pi\right)$. The symplectic eigenvalues $\lambda_j$ are defined as the square roots of the eigenvalues of $(i J G)^2$, where $J$ is the symplectic identity matrix \cite{Lemonik2016} defined as 
\begin{equation}
    J = \begin{pmatrix}
        0 & \mathbb{I} \\ -\mathbb{I} & 0
    \end{pmatrix}
\end{equation}
where $\mathbb{I}$ is the identity matrix. 
Within this framework, the $\lambda_j$ are related to the one-body spectrum $\{\omega_{j}\}$ of $\mathcal{H}_{E}$ through
\begin{equation}\label{eq:entEVs}
    \lambda_{j} = \frac{1}{2}\mathrm{coth}\Big(\frac{\omega_j}{2}\Big).
\end{equation}
The positivity of $\omega_{j}$ implies $\lambda_{j} \geq 1/2$. It is therefore convenient to introduce the entanglement occupation numbers $n_{j} = (\lambda_{j} - 1/2) \geq 0$. 
The EE can then be expressed as 
\begin{eqnarray}
     S(A|B) &=& \sum_j s_{\rm BE}(n_j),
\end{eqnarray}
where
\begin{equation}\label{eq:BE_ent}
     s_{\rm BE}(n) = (1+n)\log(1+n) - n\log n ~
\end{equation}
is the entropy of a Bose-Einstein distribution of a bosonic mode with mean population $n$.
In practice, the correlation matrix is evaluated numerically in position space, by choosing a subsystem $A$ and computing $G=\mathrm{Re}\langle \eta(\br)\eta^{\dagger}(\br')\rangle$ with the positions $(\br, \br')$ restricted to lie within $A$.

\subsection{Implementation in the $\mathcal{O}(N)$ model}

While the $\mathcal{O}(N)$ model presented in Sec.~\ref{sec:theory_field_theory} is \emph{continuous} and is naturally formulated in \emph{momentum space}, a practical numerical evaluation of the EE using the Williamson approach typically requires a \emph{discretization} in \emph{position space} to compute the correlation matrix \cite{Srednicki1993}. 
These differences necessitate constructing an appropriate lattice geometry in real space starting from the momentum-space equations of motion \eqref{eq:mode_func_eom} and \eqref{eq:effective_r}. To ensure that the EE faithfully captures the physics of the underlying continuous theory, the lattice spacing $a$ must be small enough.
\begin{figure}[h!]
  \centering
    \includegraphics[width=\linewidth]{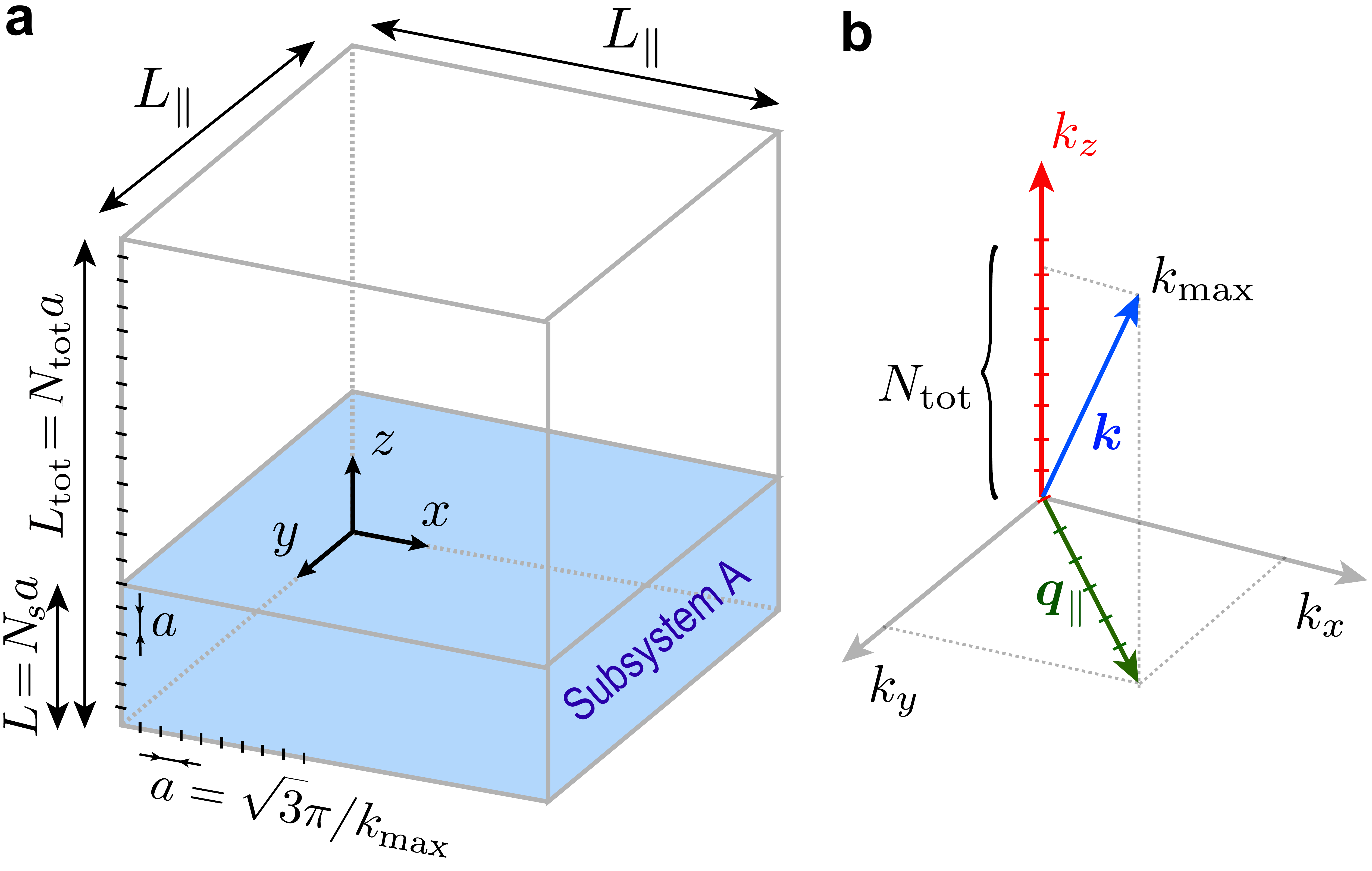}
    \caption{
     \label{fig:geometry}
    Panel (a) Spatial geometry considered in this work. We compute the entanglement entropy $S(A|B)$ in the $\mathcal{O}(N)$ model by considering   a parallelepiped subsystem A of volume $LL_\parallel^2$ with transverse extent $L_\parallel\to\infty$. 
    The system is numerically discretized, with a grid step $a$ equal in each spatial direction and fixed by the maximum radial momentum $k_\text{max}$ in panel $(b)$.
    Panel (b) Corresponding discretization of momenta in the $\mathcal{O}(N)$ model. $k_z$ is discretized in $N_\text{tot}$ steps within $[-\pi/a,\pi/a]$, with $N_\text{tot}$ setting the full system size $L_\text{tot}=N_\text{tot}a$. 
}
  \label{fig:schematic}
\end{figure}
The spatial geometry we use to evaluate the EE in the $\mathcal{O}(N)$ model is illustrated in Fig.~\ref{fig:geometry}a. The full system is modeled as a box of volume  $L_\parallel\times L_\parallel\times L_\text{tot}$ in $(x,y,z)$ space, while subsystem $A$ is taken as a parallelepiped of volume $L_\parallel^2L$. 

In Ref. \cite{Lemonik2016}, a  cubic geometry with finite $L_\parallel=L$ was employed, and the correlation matrix was subsequently spatially discretized in the full 3D volume. 
This limited the analysis to rather small subsystem sizes (with at most  $10-20$ grid steps in each direction), even with approximate analytical expressions for the correlation functions. 
To overcome this limitation, here we instead focus on the limit $L_\parallel\to\infty$ and study the EE as a function of the sole longitudinal subsystem size $L$. 
This approach allows us to probe the entanglement entropy directly in the thermodynamic limit, while still relying on exact numerical expressions for the correlation matrix.
To effectively implement this geometry starting from the $\mathcal{O}(N)$ model, we compute the correlation matrix $G$ in a \emph{mixed representation}, performing a Fourier transform with respect to the longitudinal momentum $k_z$, while staying in Fourier space in the $(x,y)$ plane: 
\begin{equation}
\label{eq:corr_matrix}
    G_{q_{\parallel}}(z-z',t)\!=\! \frac{1}{2\pi}\int_{-\pi/a}^{\pi/a} 
    \!\!\mathrm{d}k_{z} e^{ik_{z}(z- z')} \begin{pmatrix}
        G_{\phi\phi} & G_{\phi\pi} \\ 
        G_{\pi\phi}& G_{\pi\pi}
    \end{pmatrix}
\end{equation}
where the two-point $\phi-\phi$ correlation function  $G_{\phi\phi}\equiv G_{\phi\phi}(\bk,t)$ is defined in Eq.~\eqref{eq:phiphicor}, with  $\smash{k = (k_z^2 + q_\parallel^2)^{1/2}}$ (and analogous definitions for the $\phi-\pi$ and $\pi-\pi$ correlations).
In this expression, $|k_z|$ is bounded by $\smash{\pi/a\equiv k_\text{max}/\sqrt{3}}$, where $k_\text{max}$ denotes the upper cutoff used to numerically evaluate the radial integral in Eq.~\eqref{eq:effective_r}. In the simulations, we choose $k_\text{max}=10\Lambda$. This upper bound in momentum space defines the lattice spacing $a=\sqrt{3}\pi/k_\text{max}$ of the corresponding real-space grid on which we wish to calculate the EE, as illustrated in Fig.~\ref{fig:geometry}a \cite{footnote1}. With the lattice spacing thus defined, we can evaluate the correlation matrix $(G_{q_\parallel})_{m,n}=G_{q_\parallel}(|m-n|a,t)$, where the indices $m,n=1\ldots N_\text{tot}$ label the $z,z'$ coordinates, with $L_\text{tot}=N_\text{tot}a$. In our simulations the total number $N_\text{tot}$  of grid points chosen along $z$ varies from $10^2$ to $10^5$, depending on the observables computed. For more details, we refer to figure captions.

To compute the symplectic spectrum and, in turn, the EE for the subsystem $A$ of longitudinal size $L<L_\text{tot}$, we diagonalize the correlation matrix $(G_{q_\parallel})_{m,n}$ for 
$m,n=1\ldots N_s$,  with $N_s=L/a<N_\text{tot}$. 
In the mixed representation, this diagonalization is readily achieved since the correlation matrix is block-diagonal,  each block being of size $2N_s\times 2N_s$ and associated with a fixed transverse momentum $q_\parallel$. This momentum runs from $q_\parallel=0$ to $\smash{\sqrt{2/3}k_\text{max}}$ (see Fig.~\ref{fig:geometry})
and in the simulations we choose a number  $N_\parallel$ of $q_\parallel$-blocks  large enough such that the effective transverse extent $L_\parallel=N_\parallel a$ can be considered effectively infinite. Once the symplectic eigenvalues $\smash{\lambda_{q_{\parallel}}^j}$ of $G_{q_\parallel}$ are obtained (with $j=1\ldots N_s$), the EE $S(A|B)\equiv S(L|L_\text{tot}\!-\!L)$ follows as 
\begin{equation}
\label{eq:full_EE}
S(L|L_\text{tot}\!-\!L) =  \sum_{q_{\parallel}}
    S_{q_{\parallel}}(L) ,
\end{equation}
where 
\begin{equation}
\label{eq:EntEntropy}
\begin{aligned}
      S_{q_{\parallel}}(L) = \sum_j s_{\rm BE}(n_{q_{\parallel}}^j)
\end{aligned}
\end{equation}
with $\smash{n_{q_{\parallel}}^{j}=\lambda_{q_{\parallel}}^j-1/2}$, {\color{black} and $s_\text{BE}$ defined in Eq.~\eqref{eq:BE_ent} above.} We have also used the fact that $S_{\bq_{\parallel}} = S_{q_{\parallel}}$, owing to the statistical rotational invariance of the correlation functions. Finally, note that in the quench dynamics considered here, the EE depends explicitly on the post-quench time $t$. In the following subsection, we study this time evolution and focus specifically on the EE {\color{black} of a single field component per unit area, $S(L,t)\equiv S(L|L_\text{tot}-L)/(NL^{2}_{\parallel})$}.

\subsection{Application: Dynamics of entanglement entropy across the DPT}
\label{sec:entropy_dynamics}
The quench at $t = 0$ drives the system out of equilibrium and generates entanglement between its subsystems. Although the subsequent dynamics is unitary and therefore conserves the total entropy, the EE of a subsystem evolves nontrivially in time, displaying a light-cone structure associated with a propagation speed $c$ (where here $c=1$ in the chosen units). This behavior can be understood within a quasiparticle picture, in which the quench creates pairs of entangled quasiparticles that propagate ballistically through the system with velocity $c$. Whenever one quasiparticles of a pair crosses the subsystem boundary (of size $L$) while its partner remains outside, entanglement between the two regions increases, leading to a growth of the EE. As a consequence, the entanglement initially increases linearly in time within a light cone of extent $2ct$, before eventually saturating at times much longer than $t = L/(2c)$.
More specifically, because the initial state is the zero-temperature ground state of non-interacting bosons, the EE obeys an area law at short times $t \ll L/(2c)$, namely $\smash{S(L,t)\propto L^0}$ (recall that $S(L, t)$ is normalized by $\smash{L_{\parallel}^{2}}$, see previous section). At longer times, on the other hand, nonlocal correlations progressively build up across the system, causing the entanglement to grow and ultimately leading to a volume-law scaling, $S(L,t)\propto L$ for $t \gg L/(2c)$.
\begin{figure}[h!]
  \centering
{%
    \includegraphics[width=0.9\linewidth]{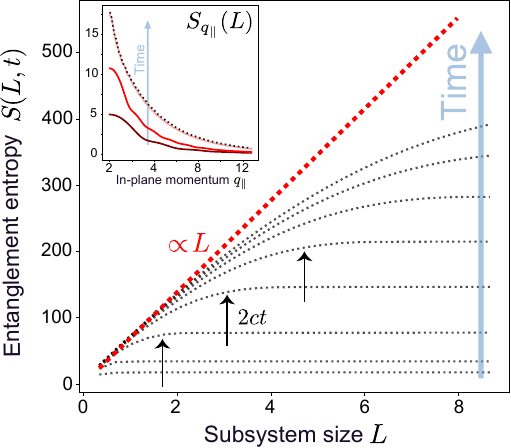}} \hfill
  \caption{
  Main panel: Entanglement entropy per unit area as a function of subsystem size at increasing times after a quench into the ordered phase ($\delta=-1$), showing the light-cone spreading of entanglement. Deep inside the light cone, $L\ll2ct$, the entropy becomes time independent and approaches $S(L,t)\propto L$ (red dotted line), displaying the volume-law scaling characteristic of entanglement spreading following the quench. Outside the light cone, $L>2ct$, where correlations have not yet propagated across the subsystem, the entropy instead obeys an area law, $\smash{S(L,t)\propto L^0}$, inherited from the initial ground state, while increasing linearly in time. Parameters: $u = 10$, $r_{i} = 100$, and $N_{\parallel} = N_{\mathrm{tot}}=200$ (corresponding to $L_{\mathrm{tot}}=L_\mathrm{tot}a \simeq 68.9$). 
 Inset: Entanglement entropy per {\color{black}$q_{\parallel}-$sector}, defined in Eq.~\eqref{eq:EntEntropy}, shown  for times $t\simeq 0.19L$, $t \simeq 0.39L$, $t \simeq 1.16L$ (solid curves from bottom to top), and $t \simeq 1.54L$ (dotted curve), for $L \simeq 5.2$.
  }
  \label{fig:area_to_volume}
\end{figure}

While these properties were previously established in the $\mathcal{O}(N)$ model in \cite{Lemonik2016} using a small subsystem and approximate expressions for correlation functions valid for ultra-deep quenches at the critical point, it is important to confirm them within an exact numerical framework. 
This is done in the main panel of Fig.~\ref{fig:area_to_volume}, which displays the EE as a function of $L$ at increasing times, computed using the procedure described in the previous sections for a representative quench below the critical point ($\delta = -1$).
At short times, $t < L/(2c)$, i.e. outside the light cone ($L > 2ct$), the EE grows linearly in time while remaining independent of $L$. This behavior reflects the area law inherited from the initial ground state. By contrast, at long times, $t > L/(2c)$, i.e. inside the light cone ($L < 2ct$), the entanglement entropy becomes nearly time independent and scales linearly with $L$, thus exhibiting a volume law. 
In the inset we also show for illustration the profile $S_{q_\parallel}$ of the EE per mode $q_{\parallel}$, defined in Eq. (\ref{eq:EntEntropy}), which slowly grows in time before eventually saturating in the long-time regime $t\gg L/(2c)$.
While Fig.~\ref{fig:area_to_volume} corresponds to a quench below the critical point, we have verified that this overall phenomenology of EE remains unchanged for quenches at and above the critical point. 
In contrast, as will be shown in Sec.~\ref{sec:entanglementgap}, sub-leading long-time corrections to the volume law of EE encode universal features of the DPT. To uncover these signatures, we now shift our focus to the entanglement properties of the $q_{\parallel}$ blocks.

\section{Low-lying dynamics of the entanglement spectrum}
\label{sec:results}

In addition to reducing computational costs, the present modeling of the EE using a subsystem infinitely extended in the $(x,y)$ plane naturally allows for a mode-resolved analysis, encapsulated in the entropy per mode, $\smash{s_\text{BE}(n_{q_\parallel}^j})$, defined in Eq.~\eqref{eq:EntEntropy}. This quantity is equivalently characterized by the occupation numbers $\smash{n^{j}_{q_{\parallel}}(t)}$,
conveniently expressed in the Bose–Einstein form \cite{Roscilde2022}
\begin{equation}
\label{eq:def_ent_disp}
  n_{q_{\parallel}}^{j}(t) = \frac{1}{\exp[{\omega^{j}_{q_{\parallel}}}(t)] - 1},
\end{equation}
which defines the entanglement dispersion relation $\smash{\omega^{j}_{q_{\parallel}}}$, with any effective temperature absorbed into $\smash{\omega^{j}_{q_{\parallel}}}$.
Physically, the entanglement dispersion supports the interpretation in terms of ``entanglement quasi-particles" that diagonalize the entanglement Hamiltonian \cite{Roscilde2022}. In the continuum limit $a\to 0$, the entanglement modes $j=1\ldots N_s$ form a continuous band. For this reason, we refer to $\smash{\omega_{q_{\parallel}}^{j}}$ as an ``entanglement band dispersion", or more simply ``entanglement spectrum".

\begin{figure}[h!]
  \centering
    {\includegraphics[width=\linewidth]{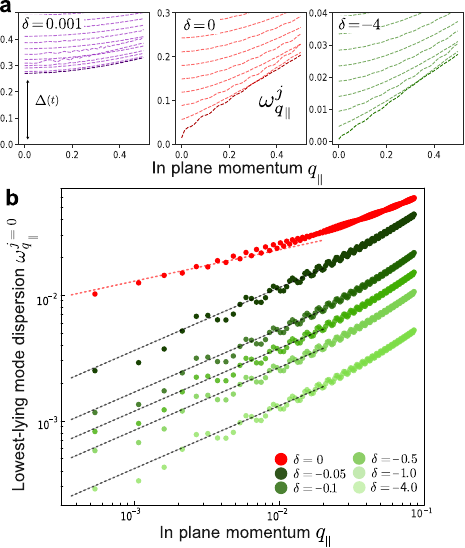}}
   \hfill
  \caption{Panel a: 
  Late-time entanglement spectrum $\smash{\omega_{q_\parallel}^j}$ as a function of $\smash{q_\parallel}$ for three different quench depths: above ($\delta=0.001$), at ($\delta=0$) and below ($\delta=-4$) the critical point of the DPT. The different dashed curves correspond to different entanglement modes $j$, with the dispersion of the lowest-lying mode $j=0$ highlighted by the darker curve. Spectra are shown at finite time $t = 200$. 
  The dispersion of the  lowest-lying zero-mode at zero momentum, $\Delta(t)\equiv\smash{\omega_{q_\parallel=0}^{j=0}}$, defines the entanglement gap.
 Panel b: Momentum dispersion of the lowest-lying mode ($j=0$) at low momentum in a log–log scale, for different quench depths $\delta$ at a fixed long time $t = 10^{3}$ {\color{black}(while $\delta = 0$ is shown at $t = 2.5\times10^{3})$. Dots represent numerical data, while black dashed curves correspond to fits of the form $\smash{\omega_{q_\parallel}^0 \propto q_{\parallel}^{1/2}}$ for quenches below the critical point. The upper red dashed curve shows a fit to $\smash{\omega_{q_\parallel}^0 \propto q_{\parallel}^{1/4}}$ for the quench at the critical point. The estimated fitting errors are of the order of $10^{-6}$ for $\delta = 0$ and $\delta = -0.05$, and of order $\leq 10^{-7}$ for larger $|\delta|$.}  
  Parameters: $N_s = 100$, $N_{\mathrm{tot}} = 5\times 10^{5}$, $u = 10$, $r_{i} = 100$. The corresponding critical mass is $r_{c} = -0.43$.}
  \label{fig:plateau_comp1}
\end{figure}

\subsection{Entanglement dispersion across the DPT}
\label{Sec:Entanglement_dispersion}

We first present the entanglement dispersion $\smash{\omega_{q_{\parallel}}^{j}}$ at long time and in the low-$q_\parallel$ regime in Fig.~\ref{fig:plateau_comp1}a, for quenches above ($\delta= 0.001$), at ($\delta= 0$), and below ($\delta= -4$) the critical point.
 {\color{black} Here and in the following, we set $r_{i} =100$, $u = 10$ and $\Lambda = \pi/2$ (see Appendix \ref{app:num_methods} for details about the numerical methods).}
The entanglement dispersion consists of a set of modes labelled by {\color{black}$j=0,1,2,\dots N_{s}$}, with the lowest-lying mode ($j = 0$) highlighted by the darker curve. According to definition~\eqref{eq:def_ent_disp}, this mode is the most highly occupied and therefore plays a central role in the long-time behavior of the total entanglement entropy~\eqref{eq:full_EE}, as will be confirmed in Sec. \ref{sec:entanglementgap}. 
Figure~\ref{fig:plateau_comp1}a first reveals a qualitatively different infrared behavior of $\smash{\omega_{q_{\parallel}}^0}$ across the dynamical phase transition: while at long time the entanglement spectrum appears gapped above the critical point (left panel of Fig.~\ref{fig:plateau_comp1}a) it asymptotically vanishes as $q_\parallel \to 0$ at and below the critical point (central and right panels).
These behaviors bear some resemblance to those of the energy dispersion $\smash{\omega_k = [k^2 + r_{\text{eff}}(t)]^{1/2}}$ across the DPT, but are nonetheless distinct. This is illustrated more clearly in Fig.~\ref{fig:plateau_comp1}b, which presents a zoom of the very low-$q_\parallel$ ($q_\parallel<0.1$) dispersion at a fixed long time, both for quenches below the critical point and for the critical quench in a log-log scale. Below the critical point, linear fits to the numerical data clearly indicate an entanglement dispersion of the form $\smash{\omega_{q_{\parallel}}^0 \propto q_\parallel^{1/2}}$ in the long-time limit. This behavior is markedly different from that of the energy dispersion, $\omega_k = [k^2 + r_{\text{eff}}(t \to \infty)]^{1/2} = k$, below the critical point.
Exactly at $\delta = 0$, by contrast, an examination of the entanglement dispersion at very low momenta instead suggests an anomalous scaling $\smash{\omega_{q_{\parallel}}^0 \propto q_\parallel^{1/4}}$ at asymptotically long times (see Fig.~\ref{fig:plateau_comp1}b). Again, such scaling does not arise for the energy dispersion at the critical point, which also behaves as $\omega_k \to k$ when $t \to \infty$.
The qualitatively different momentum scaling of the entanglement dispersion across the critical point therefore constitutes a genuine signature of the DPT at the level of entanglement properties.
To understand the physical origin of the entanglement dispersions observed in the low-energy regime, both at and below the critical point, it is useful to return to the eigenvalue equation associated with a given $(q_{\parallel},j)$ mode, which can be written as \cite{Lemonik2016}
\begin{equation}
\label{eq:EV_eq}
\begin{aligned}
(iJ\boldsymbol{G}_{q_{\parallel}})
\vec{\xi}_{q_{\parallel}}^j\! = \!\lambda_{q_{\parallel}}^j
\vec{\xi}_{q_{\parallel}}^j.
\end{aligned}
\end{equation}
Here, $\boldsymbol{G}_{q_{\parallel}}\equiv(G_{q_\parallel})_{m,n}=G_{q_\parallel}(|m-n|a,t)$ denotes the $2N_s\times2N_s$ correlation matrix, whose entries are determined by Eq. (\ref{eq:corr_matrix}), while $\smash{\lambda_{q_{\parallel}}^j}$ and $\smash{\vec{\xi}_{q_{\parallel}}^j}$ are the symplectic eigenvalue and eigenvector of the $(q_{\parallel},j)$ mode.
The low-energy structure of the entanglement dispersion is governed by the dominant $q_\parallel$-dependence of the correlation matrix in the limit $q_{\parallel} \rightarrow 0$ at late times $t/L \gg 1$. 
In particular, for the $\phi\phi$ component, Eq.~\eqref{eq:Gphiphi_lowk} implies the scaling
\begin{equation}
    G_{\phi\phi}(q_\parallel,z-z',t)\sim
    \int \frac{dk_z}{2\pi}
    \frac{e^{ik_z (z- z')}}{(k_z^2+q_\parallel^2)^{\alpha+1/2}}
    \propto q_\parallel^{-2\alpha}.
\end{equation}
We have further verified that  the other correlation functions $G_{\phi\pi}$ and $G_{\pi\pi}$ become essentially independent of $q_{\parallel}$  in the low-momentum limit. 
As a result, the full correlation matrix scales as 
\begin{equation}
\label{eq:Gqscaling}
\boldsymbol{G}_{q_{\parallel}}\sim
\begin{pmatrix}
q_\parallel^{-2\alpha} \ \mathcal{O}(1) \\ \mathcal{O}(1)\ \mathcal{O}(1)
    \end{pmatrix},
\end{equation}
where $\mathcal{O}(1)$ means a scaling of order $\smash{\sim q_{\parallel}^{0}}$. 
Applying the symplectic matrix $J$ then gives
\begin{equation}
    iJ\textbf{G}_{q_{\parallel}} \sim \begin{pmatrix}
        \mathcal{O}(1) & q_{\parallel}^{-2\alpha} \\ 
        \mathcal{O}(1) & \mathcal{O}(1)
    \end{pmatrix} \Rightarrow \lambda^{j=0}_{q_{\parallel}}\sim q_{\parallel}^{-\alpha},
\end{equation}
since the dominant momentum contribution initially located in the upper-left block of Eq.~(\ref{eq:Gqscaling}) has been transferred to the off-diagonal sector by the action of $J$. Using the relation between the entanglement spectrum and the symplectic eigenvalues, we finally obtain the low-momentum behavior of the entanglement dispersion,
\begin{equation}
    \omega_{q_\parallel}^{j=0}\propto 1/(\lambda_{q_\parallel}^{j=0})\sim
    q_\parallel^\alpha,
\end{equation}
where we recall that $\alpha=1/2$ (1/4) below (at) the critical point.
This behavior exactly  corresponds to the numerical observations of Fig. \ref{fig:plateau_comp1}b, and confirms the critical nature of the entanglement dispersion at the DPT (as well as its non-thermal character).

\begin{figure}[h!]
  \centering
{%
    \includegraphics[scale=0.45]{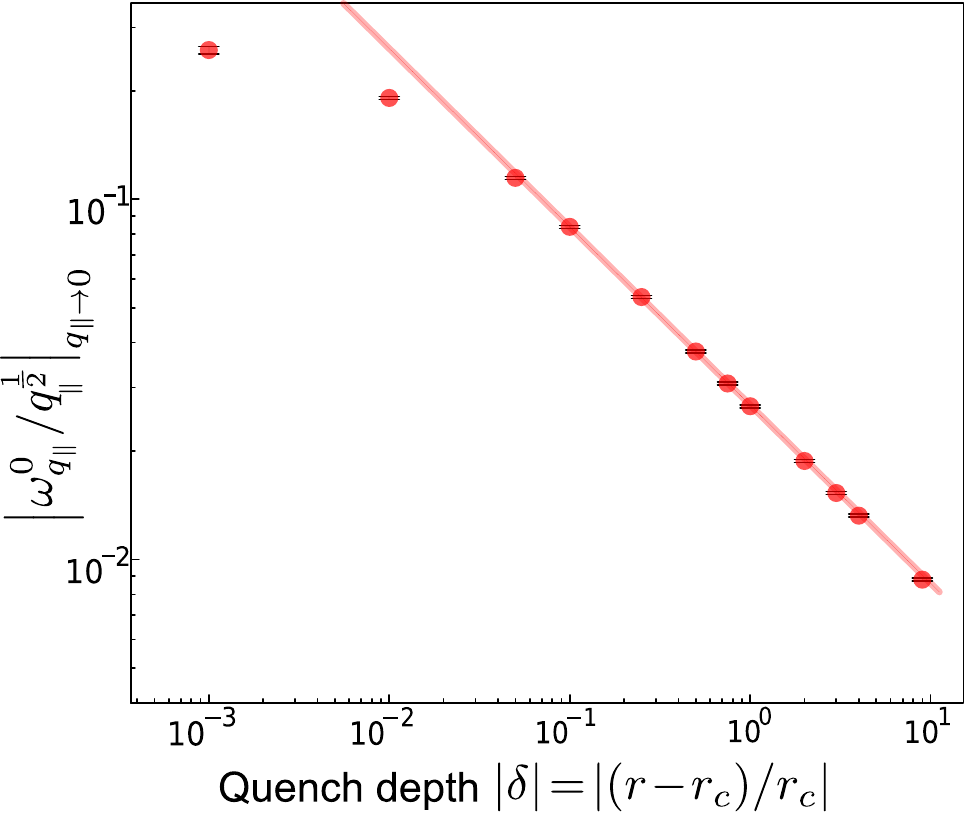}%
    \label{fig:a}} \hfill
  \caption{
  Prefactor of the entanglement dispersion, $\smash{\omega_{q_\parallel}^0/q_{\parallel}^{1/2}}$ with $q_\parallel\to0$, 
  as a function of the quench depth $|\delta|$ into the ordered phase. {\color{black}  The solid red curve corresponds to an algebraic fit of the form $ a_{1}|\delta|^{b_{1}}$ for $|\delta| \geq 0.05$, with $a_{1} = 0.027 \pm 0.006$ and $b_{1} = 0.495 \pm 0.092$.
  The numerical points are extracted from  fits of the dispersion curves in Fig.~\ref{fig:plateau_comp1}b, performed at the time $t= 10^{3}$. The error bars (black bars) are smaller than the data points, see Appendix~\ref{app:numerics} for more details on the fitting procedure.
  }
  Parameters: $r_{i} = 100$, $u = 10$, $\Lambda = \pi/2$, $N_{\mathrm{tot}} =5\times10^{5}$, and $N_s = 100$. 
 }
  \label{fig:plateau_comp2}
\end{figure}

To further confirm the critical nature of the entanglement dispersion near the DPT, it is instructive to examine how the dispersion prefactor depends on the quench parameter $\delta$ when quenching below the critical point while approaching the transition increasingly closely. To this end, Fig.~\ref{fig:plateau_comp2} shows the ratio $\smash{\omega^0_{q_\parallel}/q_\parallel^{1/2}}$ in the limit $q_\parallel \to 0$ as a function of $\delta$, extracted from the prefactors of the fitting curves displayed in Fig.~\ref{fig:plateau_comp1}b. The numerical analysis suggests a critical scaling law of the form
\begin{equation}
\label{eq:betaeff_critical}
    \frac{\omega_{q_\parallel}^0}{q_\parallel^{1/2}}
    \underset{q_\parallel\to0}{\propto} |\delta|^{-\gamma}\equiv \Big|\frac{r - r_c}{r_{c}}\Big|^{-\gamma},
\end{equation}
with $\gamma\simeq 1/2$. 
Together with the anomalous behavior of the entanglement dispersion at the critical point, this critical scaling constitutes, to the best of our knowledge, another novel feature of the DPT in the $\mathcal{O}(N)$ model at large $N$. We further note that the numerically observed ``entanglement exponent'' $\gamma$ is extremely close to the scaling exponent $\alpha = (d-2)/2 = 1/2$ governing the correlation function~\eqref{eq:self-similar}, although we have not been able to provide a simple proof that they coincide. {\color{black} Deviations from the algebraic scaling are visible for $|\delta| \leq 10^{-2}$. We believe that these may be attributed to finite-time effects, although we were unable to extend the simulations to larger times owing to numerical constraints. A detailed discussion of the fitting procedure and error estimation is provided in Appendix~\ref{app:numerics}.}
We now demonstrate that the time evolution of the entanglement gap $\smash{\omega^{0}_{q_\parallel=0}}$, which asymptotically closes below and at the critical point of the DPT, is also governed by $\alpha$.

\subsection{Dynamics of the entanglement gap}
\label{sec:entanglementgap}

Having investigated the low (but finite) $q_\parallel$  behavior of the entanglement dispersion across the DPT, we now turn to the dynamics of the ``entanglement gap'' visible in Fig. \ref{fig:plateau_comp1}a, namely the time evolution of the zero-momentum mode $\smash{\Delta(t)\equiv \omega_{q_\parallel=0}^{j=0}(t)}$. 
{\color{black} While the sensitivity of this gap to quenches to the critical point is already known, the combined numerical and analytical approach employed here allow us to considerably extend the range of times and subsystem sizes range explored in \cite{Lemonik2016}. This enables us to probe the deep-infrared properties of the ES and the late time behavior of the entanglement gap. }
In the case of equilibrium systems, the scaling of the entanglement gap with subsystem size $L$ in ordered phases was found to be responsible for logarithmic contributions to the ground-state entanglement entropy, which are sensitive to the scaling exponent \cite{Melitski2015,Frerot2015,Frerot2017}.
In the present far-from-equilibrium setting, we have seen in Sec.~\ref{sec:entropy_dynamics} that the post-quench EE typically exhibits, at leading order, a volume law within a light cone with no obvious signature of the DPT. In this section, by performing a careful scaling analysis of the entanglement gap, we derive analogous logarithmic contributions to the EE in the far-from-equilibrium context, and show that they carry clear signatures of the DPT through the scaling exponent $\alpha$. This analysis once again relies crucially on the infinitely extended nature of the subsystem, which allows us to probe the EE at asymptotically long times.

\begin{figure}[t]
\centering
\includegraphics[scale=0.42]{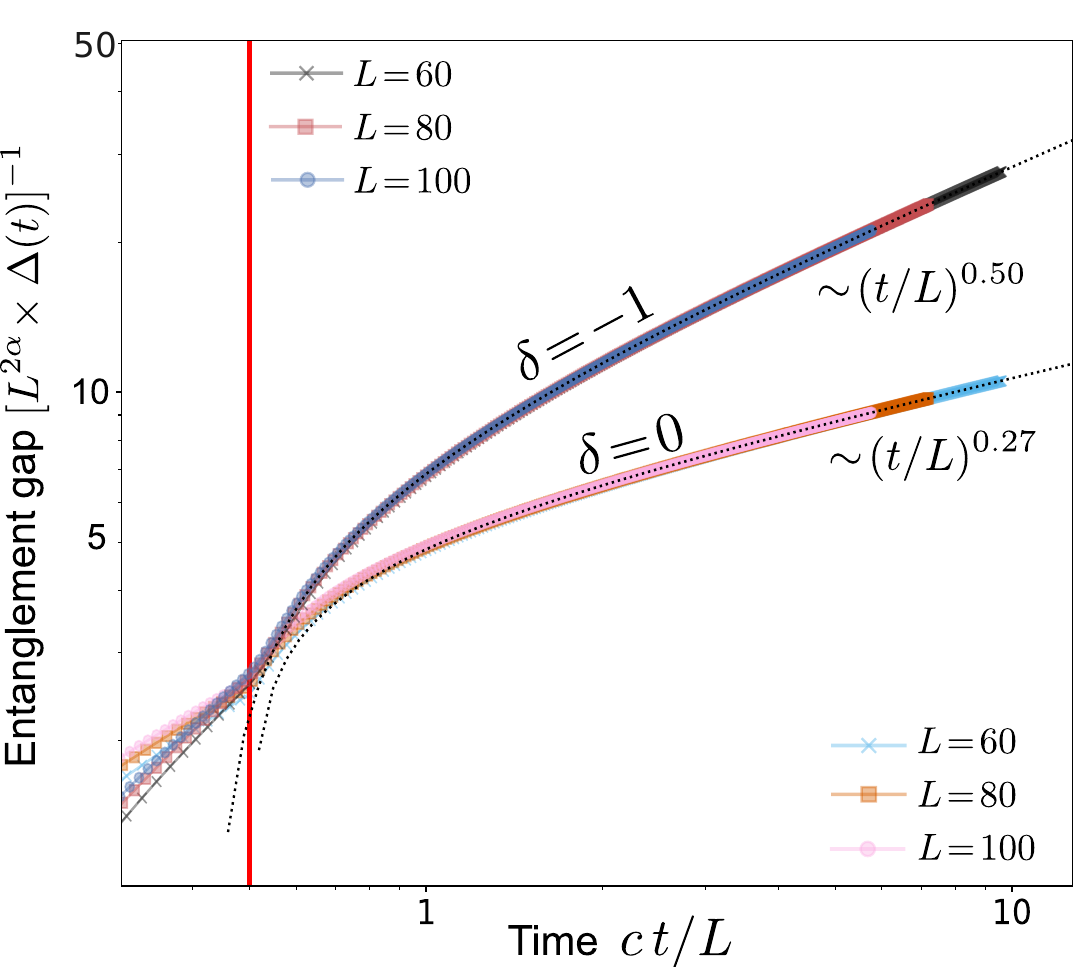}
\caption{
Inverse of the entanglement gap $\smash{\Delta(t)\equiv\omega_{q_\parallel=0}^{j=0}(t)}$ as a function of time for three subsystem sizes $L$, shown for quenches at ($\delta=0$) and below ($\delta=-1$) the critical point. In each case, the gap is rescaled by $L^{2\alpha}$ and time by $L$, using the values {\color{black}$2\alpha=0.58\pm 0.002 $ and $2\alpha=1.02 \pm 0.001$ , which are close to the scaling exponents $\alpha_c=(d-2)/4$ 
and $\alpha=(d-2)/2$ characterizing the DPT for $\delta=0$ and $\delta=-1$, respectively. These values were obtained by extracting the exponent over the range $t \in\left[190,200\right]$.} With this rescaling, data for different values of $L$ collapse onto a single curve, in agreement with the analytical predictions (see main text). The red vertical line marks the time $t = L/(2c)$, corresponding to the light-cone boundary. Beyond this point, $\Delta(t)^{-1}$ is well described by an algebraic scaling law of the form $(t-a_{\mathrm{fit}})^{b_{\mathrm{fit}}}$, with exponents {\color{black}$b_{\mathrm{fit}}\simeq 0.27 \pm 1.0\times 10^{-5} $ and $b_{\mathrm{fit}}\simeq 0.50 \pm 2\times 10^{-7}$} for $\delta=0$ and $\delta=-1$, respectively, again in close agreement with the corresponding scaling exponents.
Parameters: $r_{i} =100$, $u = 10$, $\Lambda = \pi/2$, $k_{\mathrm{max}} = 10\Lambda$ and $N_{\mathrm{tot}} = 5\times 10^{5}$.}
\label{fig:entanglement_gap}
\end{figure}

We show in Fig.~\ref{fig:entanglement_gap} the time evolution of the entanglement gap $\Delta(t)$ for quenches performed both at and below the critical point. In each case, a data collapse for different system sizes $L$ is obtained by rescaling $\Delta \to L^{2\alpha}\Delta$ and $t \to t/L$. 
This suggests the following finite-size scaling form for the gap:
\begin{equation}
\label{eq:gapscaling}
    \Delta(t)^{-1}=L^{2\alpha} W(t/L),
\end{equation}
where $\alpha$ is the scaling exponent of the DPT, $\alpha=1/2$ for quenches below the critical point, and $\alpha=\alpha_c=1/4$ for quenches at the critical point. The scaling of the entanglement gap in the latter case was  previously identified in \cite{Lemonik2016}, but for rather small subsystems and using approximate expressions for correlation functions. Our analysis demonstrates that this behavior remains robust at the exact numerical level and also extends to quenches below the critical point.
The scaling law \eqref{eq:gapscaling} can again be understood qualitatively from the simple argument based on the eigenvalue equation~\eqref{eq:EV_eq}, but now evaluated exactly at $q_\parallel=0$ \cite{Lemonik2016}. In this limit, the self-similar form~\eqref{eq:self-similar} of the field correlation function reads $G_{\phi\phi}(q_\parallel=0,k_z,t)=t^{2\alpha+1}g_{\phi\phi}(k_z t)$, where $g_{\phi\phi}\equiv f$ is a scaling function. Similarly, the mixed and momentum correlations respectively scale as  $G_{\phi\pi}=t^{2\alpha}g_{\phi\pi}(k_z t)$ and $G_{\pi\pi}=t^{2\alpha-1}g_{\pi\pi}(k_z t)$.
Introducing the rescaled fields $\bar\phi=\phi/\sqrt{t}$ and $\bar\pi=\pi\sqrt{t}$, these relations can be combined into the following scaling form for the full correlation matrix~\eqref{eq:corr_matrix} in the mixed representation~\cite{Lemonik2016}:
\begin{equation}
 G_{q_{\parallel}=0}(|m-n|a,t)\!=\!t^{2\alpha-1} 
g\Big(\frac{|m-n|a}{t}\Big),
\end{equation}
where $\boldsymbol{g}\equiv (g)_{m,n}=g(|m-n|a/t)$ is the $2N_s\times 2N_s$ rescaled correlation matrix. 
Next, we substitute this relation into the eigenvalue equation~\eqref{eq:EV_eq} evaluated at $q_\parallel = 0$, similarly to what we did for the low-energy scaling of the entanglement dispersion relation:
\begin{equation}
t^{2\alpha-1}\left(iJ\boldsymbol{g}\right)\vec{\xi}_{0}^{j} = \lambda_{0}^j\vec{\xi}_{0}^j,
\end{equation}
where we have replaced the $\smash{q_{\parallel} = 0}$ label by $0$.
We now introduce the rescaled time $\overline{t} \equiv tL$, and the rescaled  (continuum) positions $\overline{z} \equiv zL=maL$ and $\overline{z}' \equiv z'L=naL$. By virtue of the matrix relation $\smash{(\textbf{X}\vec{\xi_0})_{m} \equiv \int\mathrm{d}z' \textbf{X}(ma-z')\vec{\xi_0}(z')}$, this procedure also  rescales the eigenvalue equation $\overline{\boldsymbol{g}} = L^{-1}\boldsymbol{g}$. 
By further assuming a finite-size scaling ansatz $\lambda_{0}^{j = 0}(t) = L^{\nu} W(t/L)$ for the lowest-lying symplectic eigenvalue, the eigenvalue equation at $j=0$ can then be rewritten as 
\begin{equation}
\bar{t}^{2\alpha-1}\left(iJ\boldsymbol{\overline{g}}\right)\vec{\xi}^{j = 0}_{0} = L^{\nu -2\alpha}W(\overline{t})\vec{\xi}^{j= 0}_{0}.
\end{equation}
Requiring the right-hand side to be independent of the system size $L$ then implies $\nu = 2\alpha$, thereby recovering the scaling law~\eqref{eq:gapscaling} observed numerically.

Another interesting feature revealed by the numerical results in Fig. \ref{fig:entanglement_gap} is the long-time behavior of the scaling function $W(t/L)$. In \cite{Lemonik2016}, the analysis was performed for a finite cubic subsystem, leading to a saturation of the entanglement gap once the light-cone boundary $t=L/2$ was crossed. In the present case, by contrast, the subsystem is an infinite slab, which allows us to probe the entanglement gap in the asymptotic long-time regime. The numerical results shown in Fig. \ref{fig:entanglement_gap} suggest the asymptotic scaling form
\begin{equation}
\label{eq:Wscaling}
W(t/L)\propto \left(t/L\right)^\alpha,
\end{equation}
where $\alpha$ again denotes the scaling exponent of the DPT. This behavior can also be understood through a qualitative argument  analogous to that used in Sec. \ref{Sec:Entanglement_dispersion} to derive the scaling of the entanglement dispersion. Specifically, we use that for $q_\parallel=0$, the long-time limit of the correlation matrix is dominated by the field-field correlation (\ref{eq:self-similar}). This leads to the estimate
\begin{equation}
\label{eq:correl_scaling_def}
G_{q_{\parallel}=0}(|m\!-\!n|a,t\to\infty)\sim
\begin{pmatrix}
t^{2\alpha} & \mathcal{O}(1) \\
\mathcal{O}(1)& \mathcal{O}(1)
    \end{pmatrix}
\end{equation}
for the matrix blocks. Inserting this into the eigenvalue equation \eqref{eq:EV_eq} yields the long-time estimate for the symplectic eigenvalue at zero momentum, and in turn for the entanglement gap: $\smash{ \Delta(t\to\infty)^{-1}\simeq \lambda_{q_\parallel=0}^{j=0}(t\to\infty)\sim t^{\alpha}}$, which confirms the numerical prediction~\eqref{eq:Wscaling}. {\color{black} In Appendix \ref{app:numerics}, we present additional numerical simulations showing that both Eqs.~\eqref{eq:gapscaling} and \eqref{eq:Wscaling} are robust against a variation of the microscopic parameters of the model.}

\subsection{{\color{black}Logarithmic corrections to the entanglement entropy}}

\begin{figure}[h!]
  \includegraphics[width=0.95\linewidth]{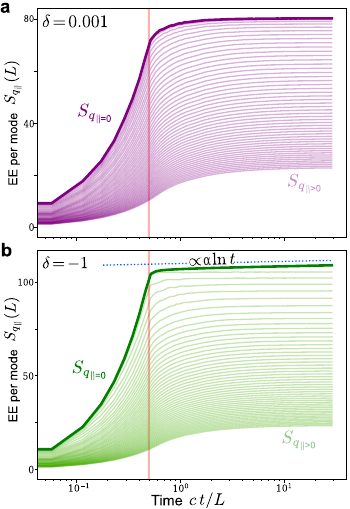}
  \caption{Entanglement entropy per $q_\parallel$ mode, $S_{q_{\parallel}}(L)$, shown for $50$ equally spaced momenta $q_{\parallel} \leq 6.34$. The bold curve corresponds to the $q_{\parallel}=0$ contribution, while the lighter curves represent the $q_{\parallel}>0$ modes in descending order.
  For quenches above the critical point (panel a), no logarithmic contribution is visible in any mode. In contrast, for quenches below the critical point (panel b), the $q_{\parallel}=0$ sector develops a $\sim \alpha \ln(t) + \mathrm{const.}$ behavior (blue dotted line) at times $t/L > 1/(2c)$, indicated by the red vertical line. This confirms the late-time scaling behavior predicted by Eq.~\eqref{eq:entropy_scaling}.
  Parameters: $u = 10$, $r_{i} = 100$, $\Lambda = \pi/2$, $N_{s} = 100$, $N_{\mathrm{tot}} = 5\times 10^{5}$ and $N_{\parallel} = 200$. }
  \label{fig:entropy_growth_log}
\end{figure}

{\color{black}
From the analysis above, we now arrive at one of the central results of this paper, namely the long-time scaling of the EE. 
Combining Eqs.~\eqref{eq:gapscaling} and~\eqref{eq:Wscaling} yields the long-time estimate $\smash{n_{q_\parallel=0}^{j=0} \sim (Lt)^\alpha}$ for the zero-mode occupation [see Eq.~\eqref{eq:def_ent_disp}]. This scaling is expected to hold inside the light cone, i.e., for times $t \gg L/(2c)$, when the EE is dominated by a volume-law contribution. In terms of the EE, this gives the following contribution from the zero mode:
\begin{equation}
\label{eq:log_s0}
\begin{aligned}
s_\text{BE}(n_{0})\!=\!(n_{0}^{0}\! +\! 1)\ln(n_{0}^{0}\! +\! 1)\! -\! n_{0}^{0}\ln(n_{0}^{0})
\sim \alpha\ln(tL).
\end{aligned}
\end{equation}
This contribution enters the full EE \eqref{eq:full_EE}, which can be rewritten as:
\begin{equation}
\label{eq:decomposedEE}
      S(L|L_\text{tot}\!-\!L) =  S_{0}(L) + 
     \int_0^\infty \frac{2\pi q_\parallel\text{d}q_{\parallel} }{(2\pi / L_{\parallel})^{2} }
     S_{q_{\parallel}}(L),
\end{equation}
where we recall that $S_{q_\parallel}(L)\equiv\sum_j s_\text{BE}(n_{q_\parallel}^j)$.  

As already observed in Fig.~\ref{fig:area_to_volume}, the dominant contribution to $S(L|L_\text{tot}\!-\!L)$ at long time is the volume law $\sim L L_\parallel^2$. 
In Appendix~\ref{app:EE_scaling}, we show that this volume law arises from the contributions of all $q_\parallel$ modes appearing in Eq.~\eqref{eq:decomposedEE}. 
Beyond the volume law, the next-to-leading contribution to $S(L|L_\text{tot}\!-\!L)$ for $t\gg L/(2c)$ is $\sim\alpha\ln t$ and originates exclusively from $S_0(L)$. This behavior is confirmed numerically in Fig.~\ref{fig:entropy_growth_log}b for $\delta=-1$ and $\alpha=1/2$. We have verified that the same logarithmic growth is also present for quenches to the critical point, $\delta=0$, with the corresponding $\alpha=1/4$.
In Appendix~\ref{app:EE_scaling}, we further show that this $\alpha\ln t$ contribution originates specifically from Eq.~\eqref{eq:log_s0}, i.e., from the $j=0$ term in $S_0$. By contrast, the $\ln L$ contribution in Eq.~\eqref{eq:log_s0} is exactly cancelled by the sum over $j>0$ in $S_0$. Finally, a detailed analysis of the $L$ dependence of both terms in Eq.~\eqref{eq:decomposedEE}, presented in Appendix~\ref{app:EE_scaling}, shows that $S_{q_\parallel}(L)$ also includes a contribution that is  independent of both $L$ and $t$, and in turn leads an area-law contribution to the EE \eqref{eq:decomposedEE}.

Altogether, at long times, $t\gg L/(2c)$, we find that the EE scales as
\begin{equation}
\label{eq:entropy_scaling}
S(L|L_{\mathrm{tot}}-L,t)
\simeq
ALL_\parallel^2 + BL_{\parallel}^2 + \alpha \ln(t) +  \dots
\end{equation}
where the coefficients $A$ and $B$ are constant in the long-time regime considered here. This asymptotic relation holds both for quenches to and below the critical point, with the corresponding value of $\alpha$. For quenches above the critical point, the zero mode remains gapped and therefore does not give rise to a logarithmic contribution, as numerically confirmed in Fig.~\ref{fig:entropy_growth_log}a.
}

Equation~(\ref{eq:entropy_scaling}) constitutes one of the main results of this work. It shows that while the time-independent volume- and area law contributions to the EE do not exhibit clear signatures of the DPT, the universal {\color{black} leading time-dependent logarithmic contribution probes the transition through its dependence on the scaling exponent $\alpha$, whose value differs for quenches below and to the critical point. 
Compared to the volume and area law contributions, it is nevertheless sub-leading as a function of $L$ and $L_{\parallel} \gg L$}. In spirit, our result is reminiscent of earlier predictions for the ground-state EE in phases with spontaneously broken continuous symmetry \cite{Melitski2015,Frerot2015,Frerot2017,wang2025}, {\color{black} where a logarithmic contribution was} likewise shown to encode universal critical properties. Here, however, the effect emerges in a genuinely far-from-equilibrium regime.


\section{Signatures of the DPT beyond the lowest lying mode}
\label{sec:ball}
In this final section, we investigate in greater detail the behavior of the lowest-lying entanglement modes at $q_\parallel=0$, as well as the qualitative differences they exhibit depending on whether the quench is performed within the disordered phase ($\delta>0$) or into the ordered phase ($\delta<0$). We examine not only the resulting changes in the entanglement spectrum, but also the spatial structure of the corresponding entanglement modes.
\begin{figure}[h!]
  \includegraphics[width=\linewidth]{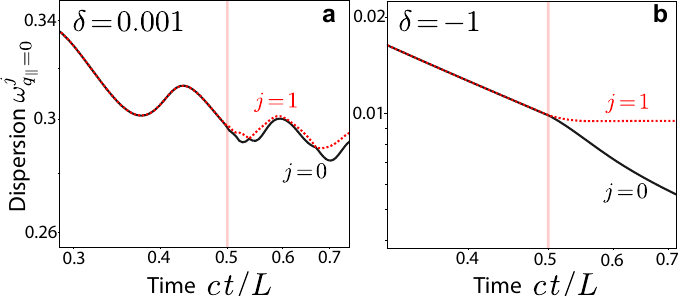}
  \caption{
  Time evolution of the dispersion at $q_{\parallel} = 0$  for the two lowest-lying entanglement modes,  $j = 0,1$. 
  Panel a: quench within the disordered phase ($\delta = 0.001$), showing an approximate degeneracy of the two eigenvalues.
  Panel b: quench into the ordered phase ($\delta = -1$), exhibiting a lifting of the degeneracy. The red vertical line marks $t = L/(2c)$. Numerical parameters are $r_{i} = 100$, $u = 10$, $\Lambda = \pi/2$,  $N_{\mathrm{tot}} = 5\times 10^{5}$, and $L = 100$.}
  \label{fig:relaxation_modes}
\end{figure}
As illustrated in Fig.~\ref{fig:relaxation_modes}, the entanglement spectrum at $q_\parallel=0$ remains two-fold degenerate up to a time $t=L/(2c)$ in both cases $\delta> 0$ or $\delta < 0$. 
This two-fold degeneracy originates from the presence of the two $A-B$ boundaries located at  $z=0$ and $z=L$, and can be understood through the following physical picture of entangled quasiparticles.
At the time of the quench ($t=0$), pairs of entangled quasiparticles are emitted throughout the system and propagate ballistically in opposite directions with velocity $c$. 
Within this picture, a point in subsystem $A$, located at a distance $l$ from a $A-B$ boundary, becomes entangled with $B$ at time $l/(2c)$, namely when it is reached by one member of a quasiparticle pair emitted midway between the point and the boundary, while the other member has crossed into $B$. 
Hence, the time $L/(2c)$ corresponds to the moment when every point in $A$  becomes entangled with regions of $B$ located on both sides of the subsystem $A$, namely at $z<0$ and at $z>L$. 
Before $t=L/(2c)$, entanglement between $A$ and $B$ 
builds up through two identical and independent processes associated with the two symmetric boundaries, whose corresponding regions are not yet mutually entangled. 
\begin{figure}[h!]
  \centering
{
    \includegraphics[width=0.9\linewidth]{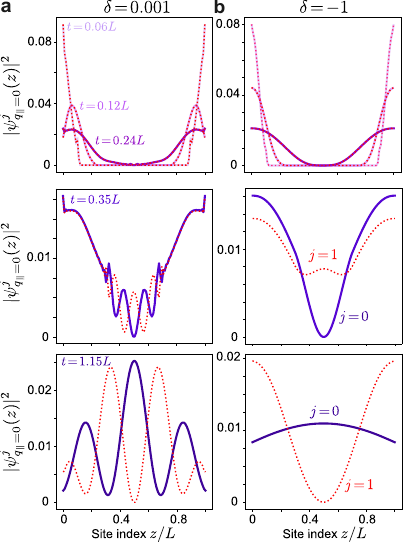}
    \label{fig:a}} \hfill
  \caption{  \label{fig:ballistic_vector}
Spatial distributions of the entanglement eigenvectors corresponding to the two lowest-lying modes, $j=0$ (solid curves) and $j=1$ (dashed curves), at $q_\parallel=0$. Left panels a: quench above the critical point $\delta=0.001$. Right panels b: quench below the critical point ($\delta=-1$).
Top row: early-time dynamics, showing ballistic entanglement fronts propagating from the subsystem boundaries toward its center with velocity $2c$. Around $t\simeq L/(4c)$, the two fronts overlap and the corresponding eigenvectors remain nearly degenerate. 
Middle row: intermediate time $t=0.35L/c$, where the eigenvalue degeneracy persists although qualitative differences between the two quench regimes already emerge. 
Bottom row: late-time regime $t>L/(2c)$. Above the critical point, the two modes remain approximately degenerate and display oscillatory spatial structures, whereas below the critical point the degeneracy is lifted and the two lowest modes exhibit well distinct, smooth spatial profiles associated with long-range correlations. Numerical parameters are $r_{i} = 100$, $u = 10$, $\Lambda = \pi/2$, $N_{\mathrm{tot}} = 5\times 10^{5}$, and  $L = 100$.}

\end{figure}
At later times, $t \ge L/(2c)$, the degeneracy between $\omega_{q_\parallel}^j$ for the $j=0,1$ modes is generally lifted at finite subsystem size. However, this lifting proceeds in qualitatively different ways above and below the critical point. For quenches within the disordered phase (Fig.~\ref{fig:relaxation_modes}a), the two entanglement modes $j=0,1$ remain nearly degenerate and oscillate around a nearby finite value, corresponding to the finite entanglement gap $\Delta$ studied in Sec.~\ref{Sec:Entanglement_dispersion} (see Fig.~\ref{fig:plateau_comp1}a). Intuitively, this behavior can be understood from the presence of a finite correlation length $\xi$: when $L \gg \xi$, regions of $B$ located on either side of $A$ remain effectively uncorrelated, so that the entanglement spectrum stays approximately two-fold degenerate. In contrast, for quenches into the ordered phase, the lifting of the degeneracy behaves qualitatively differently, see Fig.~\ref{fig:relaxation_modes}b.

While the $j=1$ entanglement mode saturates to a finite value, the $j=0$ mode continues to decay toward zero (as discussed in the previous section; see Fig.~\ref{fig:entanglement_gap}). In the entanglement spectrum, this behavior signals the emergence of a zero mode 
associated with the buildup of long-range correlations throughout the system. This
distinct dynamics
is also reflected in the spatial structure of the corresponding eigenvectors of the $iJ\textbf{G}_{q_{\parallel}}$ correlation matrices, see Eq.~\eqref{eq:EV_eq}. These are the $2N_s$ component vectors of the form
$\smash{\vec{\xi}^{j\prime}_{q_{\parallel}}\equiv (\vec{\phi}^{j\prime}_{q_{\parallel}}, \vec{\Pi}^{j\prime}_{q_{\parallel}})^{T}}$, where $\smash{\vec{\phi}^{j\prime}_{q_{\parallel}}=(\phi^{j\prime}_{q_{\parallel}}(z=0)\ldots\phi_{q_{\parallel}}^{j\prime}(z=N_sa))^{T}}$ is itself an $N_s$ component vector, since we diagonalize the entanglement Hamiltonian directly in the field basis. 
To characterize the spatial structure of the entanglement modes, we plot in Fig.~\ref{fig:ballistic_vector} the spatial distribution {\color{black}of the eigenvectors  defined as \footnote{It should be pointed out that this is by no means a unique choice to represent the evolution of the eigenvectors following the quench. This particular choice reflects the normalized eigenvectors of the $2N\times 2N$ symplectic correlation matrix $(iJC)$, however we have verified that the qualitative features presented in Fig.~\ref{fig:ballistic_vector} remain intact when plotting for example $|\phi_{q_{\parallel}}|^{2}$ or $|\Pi_{q_{\parallel}}|^{2}$ individually. Therefore, any combination of these two amplitudes will have the same qualitative features, making our choice a sufficient representation of the dynamical evolution in the modes.}
\begin{equation}
    |\psi^j_{q_\parallel}(z)|^{2} \equiv |\phi^{j\prime}_{q_{\parallel}}(z)|^{2} + |\Pi^{j\prime}_{q_{\parallel}} (z)|^{2},
\end{equation}
which is normalized $\sum_{z} |\psi^{j}_{q_\parallel}(z)|^{2} = 1$. We focus on the zero-momentum limit $q_\parallel=0$.}
As shown in the figure, for quenches both above (Fig.~\ref{fig:ballistic_vector}a) and below (Fig.~\ref{fig:ballistic_vector}b) the critical point, the $j=0$ and $j=1$ entanglement modes exhibit an entanglement front propagating from the boundaries toward the interior of subsystem $A$. This front propagates with velocity $2c$, directly reflecting the intuitive picture discussed above: a point located at a distance $l$ from the boundary of $A$ becomes entangled with $B$ after time $l/(2c)$. Consequently, the two fronts meet at the center of $A$ at time $L/(4c)$, where their spatial structures hybridize. For quenches within the disordered phase, the entanglement eigenmodes display pronounced spatio-temporal oscillations, mirroring the temporal oscillations observed in the dispersion $\omega_{q_\parallel=0}^{j=0,1}$ (Fig. \ref{fig:relaxation_modes}a). 
We find that the number of spatial oscillations increases with $L$, such that the difference between the two spatial profiles averages out in the thermodynamic limit, in close analogy with the behavior of the corresponding eigenvalues.
By contrast, for quenches into the ordered phase, the long-time spatial structure of the two modes remains clearly distinct and evolves toward smooth profiles, consistent with the emergence of a zero mode. 

\section{Conclusion}\label{sec:conclusion}
In conclusion, we have shown that the dynamical phase transition of the $\mathcal{O}(N)$ model at large $N$ leaves clear and universal fingerprints in the low-energy entanglement spectrum at late times. Specifically, we identified subleading, time-dependent logarithmic corrections to the leading volume-law contribution to the entanglement entropy, providing a direct probe of the DPT through their dependence on the scaling exponent $\alpha$.
These corrections originate from the late-time scaling properties of the lowest entanglement mode, whose dispersion and gap exhibit qualitatively distinct behaviors above, at, and below the critical point, and which we could probe  by exploiting an infinite-slab geometry for the subsystem. We also demonstrated that the entanglement spectrum contains additional information beyond the entropy itself, notably through the degeneracy lifting and the spatial structure of the entanglement eigenmodes, which reflect the buildup of long-range correlations in the ordered phase. More broadly, our results highlight the entanglement spectrum as a sensitive probe of far-from-equilibrium critical dynamics.

An interesting direction for future work would be to investigate to what extent the entanglement signatures identified here survive beyond the integrable large-$N$ limit, in particular in weakly non-integrable or fully thermalizing regimes. More broadly, it would be highly interesting to explore how similar entanglement features emerge in experimentally accessible dynamical phase transitions, such as condensation or Kosterlitz–Thouless transitions in quenched Bose gases \cite{Bornens2026}, where far-from-equilibrium scaling dynamics has recently become observable in cold-atom experiments \cite{Glidden2021, Gazo2025}. {\color{black} Finally, it would also be interesting to investigate to what extent our conclusions about DPTs extend to those witnessed in fermionic ones \cite{Jian2019, Yin2021}. A priori, as long as the resulting model can be described by an effectively quadratic density matrix, the intimate relationship between correlation matrix and EE remains intact, establishing a direct link between universal correlation properties and the entanglement spectrum.}

\section*{Acknowledgements}
The authors thank Guido Giachetti for insightful discussions. This work was supported by the French National Research Agency under Grant No. ANR-24-CE30-6695 (FUSIoN). 

\bibliography{article}

\begin{thebibliography}{65}%
\makeatletter
\providecommand \@ifxundefined [1]{%
 \@ifx{#1\undefined}
}%
\providecommand \@ifnum [1]{%
 \ifnum #1\expandafter \@firstoftwo
 \else \expandafter \@secondoftwo
 \fi
}%
\providecommand \@ifx [1]{%
 \ifx #1\expandafter \@firstoftwo
 \else \expandafter \@secondoftwo
 \fi
}%
\providecommand \natexlab [1]{#1}%
\providecommand \enquote  [1]{``#1''}%
\providecommand \bibnamefont  [1]{#1}%
\providecommand \bibfnamefont [1]{#1}%
\providecommand \citenamefont [1]{#1}%
\providecommand \href@noop [0]{\@secondoftwo}%
\providecommand \href [0]{\begingroup \@sanitize@url \@href}%
\providecommand \@href[1]{\@@startlink{#1}\@@href}%
\providecommand \@@href[1]{\endgroup#1\@@endlink}%
\providecommand \@sanitize@url [0]{\catcode `\\12\catcode `\$12\catcode `\&12\catcode `\#12\catcode `\^12\catcode `\_12\catcode `\%12\relax}%
\providecommand \@@startlink[1]{}%
\providecommand \@@endlink[0]{}%
\providecommand \url  [0]{\begingroup\@sanitize@url \@url }%
\providecommand \@url [1]{\endgroup\@href {#1}{\urlprefix }}%
\providecommand \urlprefix  [0]{URL }%
\providecommand \Eprint [0]{\href }%
\providecommand \doibase [0]{https://doi.org/}%
\providecommand \selectlanguage [0]{\@gobble}%
\providecommand \bibinfo  [0]{\@secondoftwo}%
\providecommand \bibfield  [0]{\@secondoftwo}%
\providecommand \translation [1]{[#1]}%
\providecommand \BibitemOpen [0]{}%
\providecommand \bibitemStop [0]{}%
\providecommand \bibitemNoStop [0]{.\EOS\space}%
\providecommand \EOS [0]{\spacefactor3000\relax}%
\providecommand \BibitemShut  [1]{\csname bibitem#1\endcsname}%
\let\auto@bib@innerbib\@empty
\bibitem [{\citenamefont {Hohenberg}\ and\ \citenamefont {Halperin}(1977)}]{Hohenberg1997}%
  \BibitemOpen
  \bibfield  {author} {\bibinfo {author} {\bibfnamefont {P.~C.}\ \bibnamefont {Hohenberg}}\ and\ \bibinfo {author} {\bibfnamefont {B.~I.}\ \bibnamefont {Halperin}},\ }\bibfield  {title} {\bibinfo {title} {Theory of dynamic critical phenomena},\ }\href {https://doi.org/10.1103/RevModPhys.49.435} {\bibfield  {journal} {\bibinfo  {journal} {Rev. Mod. Phys.}\ }\textbf {\bibinfo {volume} {49}},\ \bibinfo {pages} {435} (\bibinfo {year} {1977})}\BibitemShut {NoStop}%
\bibitem [{\citenamefont {Huh}\ \emph {et~al.}(2024)\citenamefont {Huh}, \citenamefont {Mukherjee}, \citenamefont {Kwon}, \citenamefont {Seo}, \citenamefont {Hur}, \citenamefont {Mistakidis}, \citenamefont {Sadeghpour},\ and\ \citenamefont {Choi}}]{Huh2024}%
  \BibitemOpen
  \bibfield  {author} {\bibinfo {author} {\bibfnamefont {S.}~\bibnamefont {Huh}}, \bibinfo {author} {\bibfnamefont {K.}~\bibnamefont {Mukherjee}}, \bibinfo {author} {\bibfnamefont {K.}~\bibnamefont {Kwon}}, \bibinfo {author} {\bibfnamefont {J.}~\bibnamefont {Seo}}, \bibinfo {author} {\bibfnamefont {J.}~\bibnamefont {Hur}}, \bibinfo {author} {\bibfnamefont {S.~I.}\ \bibnamefont {Mistakidis}}, \bibinfo {author} {\bibfnamefont {H.~R.}\ \bibnamefont {Sadeghpour}},\ and\ \bibinfo {author} {\bibfnamefont {J.-y.}\ \bibnamefont {Choi}},\ }\bibfield  {title} {\bibinfo {title} {Universality class of a spinor {B}ose–einstein condensate far from equilibrium},\ }\href {https://doi.org/10.1038/s41567-023-02339-2} {\bibfield  {journal} {\bibinfo  {journal} {Nature Physics}\ }\textbf {\bibinfo {volume} {20}},\ \bibinfo {pages} {402} (\bibinfo {year} {2024})}\BibitemShut {NoStop}%
\bibitem [{\citenamefont {Manovitz}\ \emph {et~al.}(2025)\citenamefont {Manovitz}, \citenamefont {Li}, \citenamefont {Ebadi}, \citenamefont {Samajdar}, \citenamefont {Geim}, \citenamefont {Evered}, \citenamefont {Bluvstein}, \citenamefont {Zhou}, \citenamefont {Koyluoglu}, \citenamefont {Feldmeier}, \citenamefont {Dolgirev}, \citenamefont {Maskara}, \citenamefont {Kalinowski}, \citenamefont {Sachdev}, \citenamefont {Huse}, \citenamefont {Greiner}, \citenamefont {Vuletić},\ and\ \citenamefont {Lukin}}]{Manovitz2025}%
  \BibitemOpen
  \bibfield  {author} {\bibinfo {author} {\bibfnamefont {T.}~\bibnamefont {Manovitz}}, \bibinfo {author} {\bibfnamefont {S.~H.}\ \bibnamefont {Li}}, \bibinfo {author} {\bibfnamefont {S.}~\bibnamefont {Ebadi}}, \bibinfo {author} {\bibfnamefont {R.}~\bibnamefont {Samajdar}}, \bibinfo {author} {\bibfnamefont {A.~A.}\ \bibnamefont {Geim}}, \bibinfo {author} {\bibfnamefont {S.~J.}\ \bibnamefont {Evered}}, \bibinfo {author} {\bibfnamefont {D.}~\bibnamefont {Bluvstein}}, \bibinfo {author} {\bibfnamefont {H.}~\bibnamefont {Zhou}}, \bibinfo {author} {\bibfnamefont {N.~U.}\ \bibnamefont {Koyluoglu}}, \bibinfo {author} {\bibfnamefont {J.}~\bibnamefont {Feldmeier}}, \bibinfo {author} {\bibfnamefont {P.~E.}\ \bibnamefont {Dolgirev}}, \bibinfo {author} {\bibfnamefont {N.}~\bibnamefont {Maskara}}, \bibinfo {author} {\bibfnamefont {M.}~\bibnamefont {Kalinowski}}, \bibinfo {author} {\bibfnamefont {S.}~\bibnamefont {Sachdev}}, \bibinfo {author} {\bibfnamefont {D.~A.}\ \bibnamefont {Huse}}, \bibinfo {author} {\bibfnamefont
  {M.}~\bibnamefont {Greiner}}, \bibinfo {author} {\bibfnamefont {V.}~\bibnamefont {Vuletić}},\ and\ \bibinfo {author} {\bibfnamefont {M.~D.}\ \bibnamefont {Lukin}},\ }\bibfield  {title} {\bibinfo {title} {Quantum coarsening and collective dynamics on a programmable simulator},\ }\href {https://doi.org/10.1038/s41586-024-08353-5} {\bibfield  {journal} {\bibinfo  {journal} {Nature}\ }\textbf {\bibinfo {volume} {638}},\ \bibinfo {pages} {86} (\bibinfo {year} {2025})}\BibitemShut {NoStop}%
\bibitem [{\citenamefont {Erne}\ \emph {et~al.}(2018)\citenamefont {Erne}, \citenamefont {B{\"u}cker}, \citenamefont {Gasenzer}, \citenamefont {Berges},\ and\ \citenamefont {Schmiedmayer}}]{Erne2018}%
  \BibitemOpen
  \bibfield  {author} {\bibinfo {author} {\bibfnamefont {S.}~\bibnamefont {Erne}}, \bibinfo {author} {\bibfnamefont {R.}~\bibnamefont {B{\"u}cker}}, \bibinfo {author} {\bibfnamefont {T.}~\bibnamefont {Gasenzer}}, \bibinfo {author} {\bibfnamefont {J.}~\bibnamefont {Berges}},\ and\ \bibinfo {author} {\bibfnamefont {J.}~\bibnamefont {Schmiedmayer}},\ }\bibfield  {title} {\bibinfo {title} {Universal dynamics in an isolated one-dimensional {B}ose gas far from equilibrium},\ }\href {https://doi.org/10.1038/s41586-018-0667-0} {\bibfield  {journal} {\bibinfo  {journal} {Nature}\ }\textbf {\bibinfo {volume} {563}},\ \bibinfo {pages} {225} (\bibinfo {year} {2018})}\BibitemShut {NoStop}%
\bibitem [{\citenamefont {Glidden}\ \emph {et~al.}(2021)\citenamefont {Glidden}, \citenamefont {Eigen}, \citenamefont {Dogra}, \citenamefont {Hilker}, \citenamefont {Smith},\ and\ \citenamefont {Hadzibabic}}]{Glidden2021}%
  \BibitemOpen
  \bibfield  {author} {\bibinfo {author} {\bibfnamefont {J.~A.~P.}\ \bibnamefont {Glidden}}, \bibinfo {author} {\bibfnamefont {C.}~\bibnamefont {Eigen}}, \bibinfo {author} {\bibfnamefont {L.~H.}\ \bibnamefont {Dogra}}, \bibinfo {author} {\bibfnamefont {T.~A.}\ \bibnamefont {Hilker}}, \bibinfo {author} {\bibfnamefont {R.~P.}\ \bibnamefont {Smith}},\ and\ \bibinfo {author} {\bibfnamefont {Z.}~\bibnamefont {Hadzibabic}},\ }\bibfield  {title} {\bibinfo {title} {Bidirectional dynamic scaling in an isolated {B}ose gas far from equilibrium},\ }\href {https://doi.org/10.1038/s41567-020-01114-x} {\bibfield  {journal} {\bibinfo  {journal} {Nature Physics}\ }\textbf {\bibinfo {volume} {17}},\ \bibinfo {pages} {457–461} (\bibinfo {year} {2021})}\BibitemShut {NoStop}%
\bibitem [{\citenamefont {Abuzarli}\ \emph {et~al.}(2022)\citenamefont {Abuzarli}, \citenamefont {Cherroret}, \citenamefont {Bienaim\'e},\ and\ \citenamefont {Glorieux}}]{Abuzarli2022}%
  \BibitemOpen
  \bibfield  {author} {\bibinfo {author} {\bibfnamefont {M.}~\bibnamefont {Abuzarli}}, \bibinfo {author} {\bibfnamefont {N.}~\bibnamefont {Cherroret}}, \bibinfo {author} {\bibfnamefont {T.}~\bibnamefont {Bienaim\'e}},\ and\ \bibinfo {author} {\bibfnamefont {Q.}~\bibnamefont {Glorieux}},\ }\bibfield  {title} {\bibinfo {title} {Nonequilibrium prethermal states in a two-dimensional photon fluid},\ }\href {https://doi.org/10.1103/PhysRevLett.129.100602} {\bibfield  {journal} {\bibinfo  {journal} {Phys. Rev. Lett.}\ }\textbf {\bibinfo {volume} {129}},\ \bibinfo {pages} {100602} (\bibinfo {year} {2022})}\BibitemShut {NoStop}%
\bibitem [{\citenamefont {Sunami}\ \emph {et~al.}(2023)\citenamefont {Sunami}, \citenamefont {Singh}, \citenamefont {Garrick}, \citenamefont {Beregi}, \citenamefont {Barker}, \citenamefont {Luksch}, \citenamefont {Bentine}, \citenamefont {Mathey},\ and\ \citenamefont {Foot}}]{Sunami2023}%
  \BibitemOpen
  \bibfield  {author} {\bibinfo {author} {\bibfnamefont {S.}~\bibnamefont {Sunami}}, \bibinfo {author} {\bibfnamefont {V.~P.}\ \bibnamefont {Singh}}, \bibinfo {author} {\bibfnamefont {D.}~\bibnamefont {Garrick}}, \bibinfo {author} {\bibfnamefont {A.}~\bibnamefont {Beregi}}, \bibinfo {author} {\bibfnamefont {A.~J.}\ \bibnamefont {Barker}}, \bibinfo {author} {\bibfnamefont {K.}~\bibnamefont {Luksch}}, \bibinfo {author} {\bibfnamefont {E.}~\bibnamefont {Bentine}}, \bibinfo {author} {\bibfnamefont {L.}~\bibnamefont {Mathey}},\ and\ \bibinfo {author} {\bibfnamefont {C.~J.}\ \bibnamefont {Foot}},\ }\bibfield  {title} {\bibinfo {title} {Universal scaling of the dynamic {BKT} transition in quenched {2D} {B}ose gases},\ }\href {https://www.science.org/doi/abs/10.1126/science.abq6753} {\bibfield  {journal} {\bibinfo  {journal} {Science}\ }\textbf {\bibinfo {volume} {382}},\ \bibinfo {pages} {443} (\bibinfo {year} {2023})}\BibitemShut {NoStop}%
\bibitem [{\citenamefont {Gazo}\ \emph {et~al.}(2025)\citenamefont {Gazo}, \citenamefont {Karailiev}, \citenamefont {Satoor}, \citenamefont {Eigen}, \citenamefont {Ga\l{}ka},\ and\ \citenamefont {Hadzibabic}}]{Gazo2025}%
  \BibitemOpen
  \bibfield  {author} {\bibinfo {author} {\bibfnamefont {M.}~\bibnamefont {Gazo}}, \bibinfo {author} {\bibfnamefont {A.}~\bibnamefont {Karailiev}}, \bibinfo {author} {\bibfnamefont {T.}~\bibnamefont {Satoor}}, \bibinfo {author} {\bibfnamefont {C.}~\bibnamefont {Eigen}}, \bibinfo {author} {\bibfnamefont {M.}~\bibnamefont {Ga\l{}ka}},\ and\ \bibinfo {author} {\bibfnamefont {Z.}~\bibnamefont {Hadzibabic}},\ }\bibfield  {title} {\bibinfo {title} {Universal coarsening in a homogeneous two-dimensional {B}ose gas},\ }\href {https://www.science.org/doi/abs/10.1126/science.ado3487} {\bibfield  {journal} {\bibinfo  {journal} {Science}\ }\textbf {\bibinfo {volume} {389}},\ \bibinfo {pages} {802} (\bibinfo {year} {2025})}\BibitemShut {NoStop}%
\bibitem [{\citenamefont {Berges}\ \emph {et~al.}(2008)\citenamefont {Berges}, \citenamefont {Rothkopf},\ and\ \citenamefont {Schmidt}}]{Berges2008}%
  \BibitemOpen
  \bibfield  {author} {\bibinfo {author} {\bibfnamefont {J.}~\bibnamefont {Berges}}, \bibinfo {author} {\bibfnamefont {A.}~\bibnamefont {Rothkopf}},\ and\ \bibinfo {author} {\bibfnamefont {J.}~\bibnamefont {Schmidt}},\ }\bibfield  {title} {\bibinfo {title} {Nonthermal fixed points: Effective weak coupling for strongly correlated systems far from equilibrium},\ }\href {https://doi.org/10.1103/PhysRevLett.101.041603} {\bibfield  {journal} {\bibinfo  {journal} {Phys. Rev. Lett.}\ }\textbf {\bibinfo {volume} {101}},\ \bibinfo {pages} {041603} (\bibinfo {year} {2008})}\BibitemShut {NoStop}%
\bibitem [{\citenamefont {Nowak}\ \emph {et~al.}(2012)\citenamefont {Nowak}, \citenamefont {Schole}, \citenamefont {Sexty},\ and\ \citenamefont {Gasenzer}}]{Nowak2012}%
  \BibitemOpen
  \bibfield  {author} {\bibinfo {author} {\bibfnamefont {B.}~\bibnamefont {Nowak}}, \bibinfo {author} {\bibfnamefont {J.}~\bibnamefont {Schole}}, \bibinfo {author} {\bibfnamefont {D.}~\bibnamefont {Sexty}},\ and\ \bibinfo {author} {\bibfnamefont {T.}~\bibnamefont {Gasenzer}},\ }\bibfield  {title} {\bibinfo {title} {Nonthermal fixed points, vortex statistics, and superfluid turbulence in an ultracold {B}ose gas},\ }\href {https://doi.org/10.1103/PhysRevA.85.043627} {\bibfield  {journal} {\bibinfo  {journal} {Phys. Rev. A}\ }\textbf {\bibinfo {volume} {85}},\ \bibinfo {pages} {043627} (\bibinfo {year} {2012})}\BibitemShut {NoStop}%
\bibitem [{\citenamefont {Chantesana}\ \emph {et~al.}(2019)\citenamefont {Chantesana}, \citenamefont {Orioli},\ and\ \citenamefont {Gasenzer}}]{Chantesana2019}%
  \BibitemOpen
  \bibfield  {author} {\bibinfo {author} {\bibfnamefont {I.}~\bibnamefont {Chantesana}}, \bibinfo {author} {\bibfnamefont {A.~P.}\ \bibnamefont {Orioli}},\ and\ \bibinfo {author} {\bibfnamefont {T.}~\bibnamefont {Gasenzer}},\ }\bibfield  {title} {\bibinfo {title} {Kinetic theory of nonthermal fixed points in a {B}ose gas},\ }\href {https://doi.org/10.1103/PhysRevA.99.043620} {\bibfield  {journal} {\bibinfo  {journal} {Phys. Rev. A}\ }\textbf {\bibinfo {volume} {99}},\ \bibinfo {pages} {043620} (\bibinfo {year} {2019})}\BibitemShut {NoStop}%
\bibitem [{\citenamefont {Mikheev}\ \emph {et~al.}(2019)\citenamefont {Mikheev}, \citenamefont {Schmied},\ and\ \citenamefont {Gasenzer}}]{Mikheev2019}%
  \BibitemOpen
  \bibfield  {author} {\bibinfo {author} {\bibfnamefont {A.~N.}\ \bibnamefont {Mikheev}}, \bibinfo {author} {\bibfnamefont {C.-M.}\ \bibnamefont {Schmied}},\ and\ \bibinfo {author} {\bibfnamefont {T.}~\bibnamefont {Gasenzer}},\ }\bibfield  {title} {\bibinfo {title} {Low-energy effective theory of nonthermal fixed points in a multicomponent {B}ose gas},\ }\href {https://doi.org/10.1103/PhysRevA.99.063622} {\bibfield  {journal} {\bibinfo  {journal} {Phys. Rev. A}\ }\textbf {\bibinfo {volume} {99}},\ \bibinfo {pages} {063622} (\bibinfo {year} {2019})}\BibitemShut {NoStop}%
\bibitem [{\citenamefont {Van~Regemortel}\ \emph {et~al.}(2018)\citenamefont {Van~Regemortel}, \citenamefont {Kurkjian}, \citenamefont {Wouters},\ and\ \citenamefont {Carusotto}}]{Regemortel2018}%
  \BibitemOpen
  \bibfield  {author} {\bibinfo {author} {\bibfnamefont {M.}~\bibnamefont {Van~Regemortel}}, \bibinfo {author} {\bibfnamefont {H.}~\bibnamefont {Kurkjian}}, \bibinfo {author} {\bibfnamefont {M.}~\bibnamefont {Wouters}},\ and\ \bibinfo {author} {\bibfnamefont {I.}~\bibnamefont {Carusotto}},\ }\bibfield  {title} {\bibinfo {title} {Prethermalization to thermalization crossover in a dilute {B}ose gas following an interaction ramp},\ }\href {https://doi.org/10.1103/PhysRevA.98.053612} {\bibfield  {journal} {\bibinfo  {journal} {Phys. Rev. A}\ }\textbf {\bibinfo {volume} {98}},\ \bibinfo {pages} {053612} (\bibinfo {year} {2018})}\BibitemShut {NoStop}%
\bibitem [{\citenamefont {Schmied}\ \emph {et~al.}(2019)\citenamefont {Schmied}, \citenamefont {Mikheev},\ and\ \citenamefont {Gasenzer}}]{Schmied2019}%
  \BibitemOpen
  \bibfield  {author} {\bibinfo {author} {\bibfnamefont {C.-M.}\ \bibnamefont {Schmied}}, \bibinfo {author} {\bibfnamefont {A.~N.}\ \bibnamefont {Mikheev}},\ and\ \bibinfo {author} {\bibfnamefont {T.}~\bibnamefont {Gasenzer}},\ }\bibfield  {title} {\bibinfo {title} {Prescaling in a far-from-equilibrium {B}ose gas},\ }\href {https://doi.org/10.1103/PhysRevLett.122.170404} {\bibfield  {journal} {\bibinfo  {journal} {Phys. Rev. Lett.}\ }\textbf {\bibinfo {volume} {122}},\ \bibinfo {pages} {170404} (\bibinfo {year} {2019})}\BibitemShut {NoStop}%
\bibitem [{\citenamefont {Comaron}\ \emph {et~al.}(2019)\citenamefont {Comaron}, \citenamefont {Larcher}, \citenamefont {Dalfovo},\ and\ \citenamefont {Proukakis}}]{Comaron2019}%
  \BibitemOpen
  \bibfield  {author} {\bibinfo {author} {\bibfnamefont {P.}~\bibnamefont {Comaron}}, \bibinfo {author} {\bibfnamefont {F.}~\bibnamefont {Larcher}}, \bibinfo {author} {\bibfnamefont {F.}~\bibnamefont {Dalfovo}},\ and\ \bibinfo {author} {\bibfnamefont {N.~P.}\ \bibnamefont {Proukakis}},\ }\bibfield  {title} {\bibinfo {title} {Quench dynamics of an ultracold two-dimensional {B}ose gas},\ }\href {https://doi.org/10.1103/PhysRevA.100.033618} {\bibfield  {journal} {\bibinfo  {journal} {Phys. Rev. A}\ }\textbf {\bibinfo {volume} {100}},\ \bibinfo {pages} {033618} (\bibinfo {year} {2019})}\BibitemShut {NoStop}%
\bibitem [{\citenamefont {Duval}\ and\ \citenamefont {Cherroret}(2023)}]{Duval2023}%
  \BibitemOpen
  \bibfield  {author} {\bibinfo {author} {\bibfnamefont {C.}~\bibnamefont {Duval}}\ and\ \bibinfo {author} {\bibfnamefont {N.}~\bibnamefont {Cherroret}},\ }\bibfield  {title} {\bibinfo {title} {Quantum kinetics of quenched two-dimensional {B}ose superfluids},\ }\href {http://dx.doi.org/10.1103/PhysRevA.107.043305} {\bibfield  {journal} {\bibinfo  {journal} {Phys. Rev. A}\ }\textbf {\bibinfo {volume} {107}} (\bibinfo {year} {2023})}\BibitemShut {NoStop}%
\bibitem [{\citenamefont {Gliott}\ \emph {et~al.}(2024)\citenamefont {Gliott}, \citenamefont {Ran\ifmmode~\mbox{\c{c}}\else \c{c}\fi{}on},\ and\ \citenamefont {Cherroret}}]{Gliott2024}%
  \BibitemOpen
  \bibfield  {author} {\bibinfo {author} {\bibfnamefont {E.}~\bibnamefont {Gliott}}, \bibinfo {author} {\bibfnamefont {A.}~\bibnamefont {Ran\ifmmode~\mbox{\c{c}}\else \c{c}\fi{}on}},\ and\ \bibinfo {author} {\bibfnamefont {N.}~\bibnamefont {Cherroret}},\ }\bibfield  {title} {\bibinfo {title} {From inverse-cascade to subdiffusive dynamic scaling in driven disordered {B}ose fluids},\ }\href {https://doi.org/10.1103/PhysRevLett.133.233403} {\bibfield  {journal} {\bibinfo  {journal} {Phys. Rev. Lett.}\ }\textbf {\bibinfo {volume} {133}},\ \bibinfo {pages} {233403} (\bibinfo {year} {2024})}\BibitemShut {NoStop}%
\bibitem [{\citenamefont {Duval}\ and\ \citenamefont {Cherroret}(2025)}]{Duval2025}%
  \BibitemOpen
  \bibfield  {author} {\bibinfo {author} {\bibfnamefont {C.}~\bibnamefont {Duval}}\ and\ \bibinfo {author} {\bibfnamefont {N.}~\bibnamefont {Cherroret}},\ }\bibfield  {title} {\bibinfo {title} {Anomalous landau damping and algebraic thermalization in two-dimensional superfluids far from equilibrium},\ }\href {https://doi.org/10.1103/PhysRevA.111.L021301} {\bibfield  {journal} {\bibinfo  {journal} {Phys. Rev. A}\ }\textbf {\bibinfo {volume} {111}},\ \bibinfo {pages} {L021301} (\bibinfo {year} {2025})}\BibitemShut {NoStop}%
\bibitem [{\citenamefont {Alilou}\ \emph {et~al.}(2025)\citenamefont {Alilou}, \citenamefont {Duval}, \citenamefont {Pozo},\ and\ \citenamefont {Cherroret}}]{Alilou2025}%
  \BibitemOpen
  \bibfield  {author} {\bibinfo {author} {\bibfnamefont {B.}~\bibnamefont {Alilou}}, \bibinfo {author} {\bibfnamefont {C.}~\bibnamefont {Duval}}, \bibinfo {author} {\bibfnamefont {F.~D.}\ \bibnamefont {Pozo}},\ and\ \bibinfo {author} {\bibfnamefont {N.}~\bibnamefont {Cherroret}},\ }\href {https://arxiv.org/abs/2511.14861} {\bibinfo {title} {Phonon scattering from spatial relaxation of one-dimensional {B}ose gases}} (\bibinfo {year} {2025}),\ \Eprint {https://arxiv.org/abs/2511.14861} {arXiv:2511.14861 [cond-mat.quant-gas]} \BibitemShut {NoStop}%
\bibitem [{\citenamefont {Martone}\ \emph {et~al.}(2018)\citenamefont {Martone}, \citenamefont {Larr\'e}, \citenamefont {Fabbri},\ and\ \citenamefont {Pavloff}}]{paper:quench_prot}%
  \BibitemOpen
  \bibfield  {author} {\bibinfo {author} {\bibfnamefont {G.~I.}\ \bibnamefont {Martone}}, \bibinfo {author} {\bibfnamefont {P.-E.}\ \bibnamefont {Larr\'e}}, \bibinfo {author} {\bibfnamefont {A.}~\bibnamefont {Fabbri}},\ and\ \bibinfo {author} {\bibfnamefont {N.}~\bibnamefont {Pavloff}},\ }\bibfield  {title} {\bibinfo {title} {Momentum distribution and coherence of a weakly interacting {B}ose gas after a quench},\ }\href {https://doi.org/10.1103/PhysRevA.98.063617} {\bibfield  {journal} {\bibinfo  {journal} {Phys. Rev. A}\ }\textbf {\bibinfo {volume} {98}},\ \bibinfo {pages} {063617} (\bibinfo {year} {2018})}\BibitemShut {NoStop}%
\bibitem [{\citenamefont {Bray}(1993)}]{Bray1993}%
  \BibitemOpen
  \bibfield  {author} {\bibinfo {author} {\bibfnamefont {A.}~\bibnamefont {Bray}},\ }\bibfield  {title} {\bibinfo {title} {Theory of phase ordering kinetics},\ }\href {https://doi.org/https://doi.org/10.1016/0378-4371(93)90338-5} {\bibfield  {journal} {\bibinfo  {journal} {Physica A: Statistical Mechanics and its Applications}\ }\textbf {\bibinfo {volume} {194}},\ \bibinfo {pages} {41} (\bibinfo {year} {1993})}\BibitemShut {NoStop}%
\bibitem [{\citenamefont {Berthier}\ \emph {et~al.}(2001)\citenamefont {Berthier}, \citenamefont {Holdsworth},\ and\ \citenamefont {Sellitto}}]{Berthier2001}%
  \BibitemOpen
  \bibfield  {author} {\bibinfo {author} {\bibfnamefont {L.}~\bibnamefont {Berthier}}, \bibinfo {author} {\bibfnamefont {P.~C.~W.}\ \bibnamefont {Holdsworth}},\ and\ \bibinfo {author} {\bibfnamefont {M.}~\bibnamefont {Sellitto}},\ }\bibfield  {title} {\bibinfo {title} {Nonequilibrium critical dynamics of the two-dimensional {XY} model},\ }\href {https://doi.org/10.1088/0305-4470/34/9/301} {\bibfield  {journal} {\bibinfo  {journal} {Journal of Physics A: Mathematical and General}\ }\textbf {\bibinfo {volume} {34}},\ \bibinfo {pages} {1805} (\bibinfo {year} {2001})}\BibitemShut {NoStop}%
\bibitem [{\citenamefont {Cugliandolo}(2015)}]{Cugliandolo2015}%
  \BibitemOpen
  \bibfield  {author} {\bibinfo {author} {\bibfnamefont {L.~F.}\ \bibnamefont {Cugliandolo}},\ }\bibfield  {title} {\bibinfo {title} {Coarsening phenomena},\ }\href {https://doi.org/https://doi.org/10.1016/j.crhy.2015.02.005} {\bibfield  {journal} {\bibinfo  {journal} {Comptes Rendus Physique}\ }\textbf {\bibinfo {volume} {16}},\ \bibinfo {pages} {257} (\bibinfo {year} {2015})},\ \bibinfo {note} {coarsening dynamics / Dynamique de coarsening}\BibitemShut {NoStop}%
\bibitem [{\citenamefont {Gliott}\ \emph {et~al.}(2025)\citenamefont {Gliott}, \citenamefont {Piekarski},\ and\ \citenamefont {Cherroret}}]{Gliott2025}%
  \BibitemOpen
  \bibfield  {author} {\bibinfo {author} {\bibfnamefont {E.}~\bibnamefont {Gliott}}, \bibinfo {author} {\bibfnamefont {C.}~\bibnamefont {Piekarski}},\ and\ \bibinfo {author} {\bibfnamefont {N.}~\bibnamefont {Cherroret}},\ }\bibfield  {title} {\bibinfo {title} {Coarsening of binary {B}ose superfluids: An effective theory},\ }\href {https://doi.org/10.1103/2rqw-prly} {\bibfield  {journal} {\bibinfo  {journal} {Phys. Rev. Res.}\ }\textbf {\bibinfo {volume} {7}},\ \bibinfo {pages} {033189} (\bibinfo {year} {2025})}\BibitemShut {NoStop}%
\bibitem [{\citenamefont {Balducci}\ \emph {et~al.}(2026)\citenamefont {Balducci}, \citenamefont {Chandran},\ and\ \citenamefont {Moessner}}]{Balducci2026}%
  \BibitemOpen
  \bibfield  {author} {\bibinfo {author} {\bibfnamefont {F.}~\bibnamefont {Balducci}}, \bibinfo {author} {\bibfnamefont {A.}~\bibnamefont {Chandran}},\ and\ \bibinfo {author} {\bibfnamefont {R.}~\bibnamefont {Moessner}},\ }\bibfield  {title} {\bibinfo {title} {Symmetry rebreaking in an effective theory of quantum coarsening},\ }\href {https://doi.org/10.1103/q1lm-99z4} {\bibfield  {journal} {\bibinfo  {journal} {Phys. Rev. Lett.}\ }\textbf {\bibinfo {volume} {136}},\ \bibinfo {pages} {020402} (\bibinfo {year} {2026})}\BibitemShut {NoStop}%
\bibitem [{\citenamefont {Moshe}\ and\ \citenamefont {Zinn-Justin}(2003)}]{Moshe2003}%
  \BibitemOpen
  \bibfield  {author} {\bibinfo {author} {\bibfnamefont {M.}~\bibnamefont {Moshe}}\ and\ \bibinfo {author} {\bibfnamefont {J.}~\bibnamefont {Zinn-Justin}},\ }\bibfield  {title} {\bibinfo {title} {Quantum field theory in the large {N} limit: a review},\ }\href {https://doi.org/https://doi.org/10.1016/S0370-1573(03)00263-1} {\bibfield  {journal} {\bibinfo  {journal} {Physics Reports}\ }\textbf {\bibinfo {volume} {385}},\ \bibinfo {pages} {69} (\bibinfo {year} {2003})}\BibitemShut {NoStop}%
\bibitem [{\citenamefont {Chandran}\ \emph {et~al.}(2013)\citenamefont {Chandran}, \citenamefont {Nanduri}, \citenamefont {Gubser},\ and\ \citenamefont {Sondhi}}]{Chandran2013}%
  \BibitemOpen
  \bibfield  {author} {\bibinfo {author} {\bibfnamefont {A.}~\bibnamefont {Chandran}}, \bibinfo {author} {\bibfnamefont {A.}~\bibnamefont {Nanduri}}, \bibinfo {author} {\bibfnamefont {S.~S.}\ \bibnamefont {Gubser}},\ and\ \bibinfo {author} {\bibfnamefont {S.~L.}\ \bibnamefont {Sondhi}},\ }\bibfield  {title} {\bibinfo {title} {Equilibration and coarsening in the quantum ${O(N)}$ model at infinite ${N}$},\ }\href {https://doi.org/10.1103/PhysRevB.88.024306} {\bibfield  {journal} {\bibinfo  {journal} {Phys. Rev. B}\ }\textbf {\bibinfo {volume} {88}},\ \bibinfo {pages} {024306} (\bibinfo {year} {2013})}\BibitemShut {NoStop}%
\bibitem [{\citenamefont {Sciolla}\ and\ \citenamefont {Biroli}(2013)}]{Sciolla2013}%
  \BibitemOpen
  \bibfield  {author} {\bibinfo {author} {\bibfnamefont {B.}~\bibnamefont {Sciolla}}\ and\ \bibinfo {author} {\bibfnamefont {G.}~\bibnamefont {Biroli}},\ }\bibfield  {title} {\bibinfo {title} {Quantum quenches, dynamical transitions, and off-equilibrium quantum criticality},\ }\href {https://doi.org/10.1103/PhysRevB.88.201110} {\bibfield  {journal} {\bibinfo  {journal} {Phys. Rev. B}\ }\textbf {\bibinfo {volume} {88}},\ \bibinfo {pages} {201110} (\bibinfo {year} {2013})}\BibitemShut {NoStop}%
\bibitem [{\citenamefont {Maraga}\ \emph {et~al.}(2015)\citenamefont {Maraga}, \citenamefont {Chiocchetta}, \citenamefont {Mitra},\ and\ \citenamefont {Gambassi}}]{Maraga2015}%
  \BibitemOpen
  \bibfield  {author} {\bibinfo {author} {\bibfnamefont {A.}~\bibnamefont {Maraga}}, \bibinfo {author} {\bibfnamefont {A.}~\bibnamefont {Chiocchetta}}, \bibinfo {author} {\bibfnamefont {A.}~\bibnamefont {Mitra}},\ and\ \bibinfo {author} {\bibfnamefont {A.}~\bibnamefont {Gambassi}},\ }\bibfield  {title} {\bibinfo {title} {Aging and coarsening in isolated quantum systems after a quench: Exact results for the quantum ${O}({N})$ model with ${N}$ $\ensuremath{\rightarrow}$ $\ensuremath{\infty}$},\ }\href {https://doi.org/10.1103/PhysRevE.92.042151} {\bibfield  {journal} {\bibinfo  {journal} {Phys. Rev. E}\ }\textbf {\bibinfo {volume} {92}},\ \bibinfo {pages} {042151} (\bibinfo {year} {2015})}\BibitemShut {NoStop}%
\bibitem [{\citenamefont {Smacchia}\ \emph {et~al.}(2015)\citenamefont {Smacchia}, \citenamefont {Knap}, \citenamefont {Demler},\ and\ \citenamefont {Silva}}]{Smacchia2015}%
  \BibitemOpen
  \bibfield  {author} {\bibinfo {author} {\bibfnamefont {P.}~\bibnamefont {Smacchia}}, \bibinfo {author} {\bibfnamefont {M.}~\bibnamefont {Knap}}, \bibinfo {author} {\bibfnamefont {E.}~\bibnamefont {Demler}},\ and\ \bibinfo {author} {\bibfnamefont {A.}~\bibnamefont {Silva}},\ }\bibfield  {title} {\bibinfo {title} {Exploring dynamical phase transitions and prethermalization with quantum noise of excitations},\ }\href {https://doi.org/10.1103/PhysRevB.91.205136} {\bibfield  {journal} {\bibinfo  {journal} {Phys. Rev. B}\ }\textbf {\bibinfo {volume} {91}},\ \bibinfo {pages} {205136} (\bibinfo {year} {2015})}\BibitemShut {NoStop}%
\bibitem [{\citenamefont {Chiocchetta}\ \emph {et~al.}(2017)\citenamefont {Chiocchetta}, \citenamefont {Gambassi}, \citenamefont {Diehl},\ and\ \citenamefont {Marino}}]{Diehl2017}%
  \BibitemOpen
  \bibfield  {author} {\bibinfo {author} {\bibfnamefont {A.}~\bibnamefont {Chiocchetta}}, \bibinfo {author} {\bibfnamefont {A.}~\bibnamefont {Gambassi}}, \bibinfo {author} {\bibfnamefont {S.}~\bibnamefont {Diehl}},\ and\ \bibinfo {author} {\bibfnamefont {J.}~\bibnamefont {Marino}},\ }\bibfield  {title} {\bibinfo {title} {Dynamical crossovers in prethermal critical states},\ }\href {https://link.aps.org/doi/10.1103/PhysRevLett.118.135701} {\bibfield  {journal} {\bibinfo  {journal} {Phys. Rev. Lett.}\ }\textbf {\bibinfo {volume} {118}},\ \bibinfo {pages} {135701} (\bibinfo {year} {2017})}\BibitemShut {NoStop}%
\bibitem [{\citenamefont {Halimeh}\ and\ \citenamefont {Maghrebi}(2021)}]{Halimeh2021}%
  \BibitemOpen
  \bibfield  {author} {\bibinfo {author} {\bibfnamefont {J.~C.}\ \bibnamefont {Halimeh}}\ and\ \bibinfo {author} {\bibfnamefont {M.~F.}\ \bibnamefont {Maghrebi}},\ }\bibfield  {title} {\bibinfo {title} {Quantum aging and dynamical universality in the long-range ${O(N}\ensuremath{\rightarrow}\ensuremath{\infty})$ model},\ }\href {https://doi.org/10.1103/PhysRevE.103.052142} {\bibfield  {journal} {\bibinfo  {journal} {Phys. Rev. E}\ }\textbf {\bibinfo {volume} {103}},\ \bibinfo {pages} {052142} (\bibinfo {year} {2021})}\BibitemShut {NoStop}%
\bibitem [{\citenamefont {Cherroret}(2024)}]{Cherroret2024}%
  \BibitemOpen
  \bibfield  {author} {\bibinfo {author} {\bibfnamefont {N.}~\bibnamefont {Cherroret}},\ }\bibfield  {title} {\bibinfo {title} {Dynamical phase transition of light in time-varying nonlinear dispersive media},\ }\href {https://doi.org/10.1103/PhysRevA.109.013519} {\bibfield  {journal} {\bibinfo  {journal} {Phys. Rev. A}\ }\textbf {\bibinfo {volume} {109}},\ \bibinfo {pages} {013519} (\bibinfo {year} {2024})}\BibitemShut {NoStop}%
\bibitem [{\citenamefont {Giachetti}\ \emph {et~al.}(2025)\citenamefont {Giachetti}, \citenamefont {Solfanelli},\ and\ \citenamefont {Defenu}}]{Giachetti2025}%
  \BibitemOpen
  \bibfield  {author} {\bibinfo {author} {\bibfnamefont {G.}~\bibnamefont {Giachetti}}, \bibinfo {author} {\bibfnamefont {A.}~\bibnamefont {Solfanelli}},\ and\ \bibinfo {author} {\bibfnamefont {N.}~\bibnamefont {Defenu}},\ }\href {https://arxiv.org/abs/2511.08687} {\bibinfo {title} {Universality and weak-ergodicity breaking in quantum quenches}} (\bibinfo {year} {2025}),\ \Eprint {https://arxiv.org/abs/2511.08687} {arXiv:2511.08687 [cond-mat.stat-mech]} \BibitemShut {NoStop}%
\bibitem [{\citenamefont {Laflorencie}(2016)}]{Laflorencie2016}%
  \BibitemOpen
  \bibfield  {author} {\bibinfo {author} {\bibfnamefont {N.}~\bibnamefont {Laflorencie}},\ }\bibfield  {title} {\bibinfo {title} {Quantum entanglement in condensed matter systems},\ }\href {https://doi.org/10.1016/j.physrep.2016.06.008} {\bibfield  {journal} {\bibinfo  {journal} {Physics Reports}\ }\textbf {\bibinfo {volume} {646}},\ \bibinfo {pages} {1–59} (\bibinfo {year} {2016})}\BibitemShut {NoStop}%
\bibitem [{\citenamefont {Turner}\ \emph {et~al.}(2011)\citenamefont {Turner}, \citenamefont {Pollmann},\ and\ \citenamefont {Berg}}]{Turner2011}%
  \BibitemOpen
  \bibfield  {author} {\bibinfo {author} {\bibfnamefont {A.~M.}\ \bibnamefont {Turner}}, \bibinfo {author} {\bibfnamefont {F.}~\bibnamefont {Pollmann}},\ and\ \bibinfo {author} {\bibfnamefont {E.}~\bibnamefont {Berg}},\ }\bibfield  {title} {\bibinfo {title} {Topological phases of one-dimensional fermions: An entanglement point of view},\ }\href {http://dx.doi.org/10.1103/PhysRevB.83.075102} {\bibfield  {journal} {\bibinfo  {journal} {Phys. Rev. B}\ }\textbf {\bibinfo {volume} {83}} (\bibinfo {year} {2011})}\BibitemShut {NoStop}%
\bibitem [{\citenamefont {Metlitski}\ \emph {et~al.}(2009)\citenamefont {Metlitski}, \citenamefont {Fuertes},\ and\ \citenamefont {Sachdev}}]{Metlitski2009}%
  \BibitemOpen
  \bibfield  {author} {\bibinfo {author} {\bibfnamefont {M.~A.}\ \bibnamefont {Metlitski}}, \bibinfo {author} {\bibfnamefont {C.~A.}\ \bibnamefont {Fuertes}},\ and\ \bibinfo {author} {\bibfnamefont {S.}~\bibnamefont {Sachdev}},\ }\bibfield  {title} {\bibinfo {title} {Entanglement entropy in the ${O(N)}$ model},\ }\href {https://link.aps.org/doi/10.1103/PhysRevB.80.115122} {\bibfield  {journal} {\bibinfo  {journal} {Phys. Rev. B}\ }\textbf {\bibinfo {volume} {80}},\ \bibinfo {pages} {115122} (\bibinfo {year} {2009})}\BibitemShut {NoStop}%
\bibitem [{\citenamefont {Casini}\ and\ \citenamefont {Huerta}(2012)}]{Casini2012}%
  \BibitemOpen
  \bibfield  {author} {\bibinfo {author} {\bibfnamefont {H.}~\bibnamefont {Casini}}\ and\ \bibinfo {author} {\bibfnamefont {M.}~\bibnamefont {Huerta}},\ }\bibfield  {title} {\bibinfo {title} {Renormalization group running of the entanglement entropy of a circle},\ }\href {https://doi.org/10.1103/PhysRevD.85.125016} {\bibfield  {journal} {\bibinfo  {journal} {Phys. Rev. D}\ }\textbf {\bibinfo {volume} {85}},\ \bibinfo {pages} {125016} (\bibinfo {year} {2012})}\BibitemShut {NoStop}%
\bibitem [{\citenamefont {Kallin}\ \emph {et~al.}(2013)\citenamefont {Kallin}, \citenamefont {Hyatt}, \citenamefont {Singh},\ and\ \citenamefont {Melko}}]{Kallin2013}%
  \BibitemOpen
  \bibfield  {author} {\bibinfo {author} {\bibfnamefont {A.~B.}\ \bibnamefont {Kallin}}, \bibinfo {author} {\bibfnamefont {K.}~\bibnamefont {Hyatt}}, \bibinfo {author} {\bibfnamefont {R.~R.~P.}\ \bibnamefont {Singh}},\ and\ \bibinfo {author} {\bibfnamefont {R.~G.}\ \bibnamefont {Melko}},\ }\bibfield  {title} {\bibinfo {title} {Entanglement at a two-dimensional quantum critical point: A numerical linked-cluster expansion study},\ }\href {https://doi.org/10.1103/PhysRevLett.110.135702} {\bibfield  {journal} {\bibinfo  {journal} {Phys. Rev. Lett.}\ }\textbf {\bibinfo {volume} {110}},\ \bibinfo {pages} {135702} (\bibinfo {year} {2013})}\BibitemShut {NoStop}%
\bibitem [{\citenamefont {Helmes}\ and\ \citenamefont {Wessel}(2014)}]{Helmes2014}%
  \BibitemOpen
  \bibfield  {author} {\bibinfo {author} {\bibfnamefont {J.}~\bibnamefont {Helmes}}\ and\ \bibinfo {author} {\bibfnamefont {S.}~\bibnamefont {Wessel}},\ }\bibfield  {title} {\bibinfo {title} {Entanglement entropy scaling in the bilayer heisenberg spin system},\ }\href {https://doi.org/10.1103/PhysRevB.89.245120} {\bibfield  {journal} {\bibinfo  {journal} {Phys. Rev. B}\ }\textbf {\bibinfo {volume} {89}},\ \bibinfo {pages} {245120} (\bibinfo {year} {2014})}\BibitemShut {NoStop}%
\bibitem [{\citenamefont {Fr\'erot}\ and\ \citenamefont {Roscilde}(2016)}]{Frerot2016}%
  \BibitemOpen
  \bibfield  {author} {\bibinfo {author} {\bibfnamefont {I.}~\bibnamefont {Fr\'erot}}\ and\ \bibinfo {author} {\bibfnamefont {T.}~\bibnamefont {Roscilde}},\ }\bibfield  {title} {\bibinfo {title} {Entanglement entropy across the superfluid-insulator transition: A signature of bosonic criticality},\ }\href {https://doi.org/10.1103/PhysRevLett.116.190401} {\bibfield  {journal} {\bibinfo  {journal} {Phys. Rev. Lett.}\ }\textbf {\bibinfo {volume} {116}},\ \bibinfo {pages} {190401} (\bibinfo {year} {2016})}\BibitemShut {NoStop}%
\bibitem [{\citenamefont {Eisler}\ and\ \citenamefont {Peschel}(2008)}]{Peschel2008}%
  \BibitemOpen
  \bibfield  {author} {\bibinfo {author} {\bibfnamefont {V.}~\bibnamefont {Eisler}}\ and\ \bibinfo {author} {\bibfnamefont {I.}~\bibnamefont {Peschel}},\ }\bibfield  {title} {\bibinfo {title} {Entanglement in a periodic quench},\ }\href {https://doi.org/10.1002/andp.20085200605} {\bibfield  {journal} {\bibinfo  {journal} {Annalen der Physik}\ }\textbf {\bibinfo {volume} {520}},\ \bibinfo {pages} {410–423} (\bibinfo {year} {2008})}\BibitemShut {NoStop}%
\bibitem [{\citenamefont {Calabrese}\ and\ \citenamefont {Cardy}(2007)}]{Calabrese2007}%
  \BibitemOpen
  \bibfield  {author} {\bibinfo {author} {\bibfnamefont {P.}~\bibnamefont {Calabrese}}\ and\ \bibinfo {author} {\bibfnamefont {J.}~\bibnamefont {Cardy}},\ }\bibfield  {title} {\bibinfo {title} {Entanglement and correlation functions following a local quench: a conformal field theory approach},\ }\href {https://doi.org/10.1088/1742-5468/2007/10/p10004} {\bibfield  {journal} {\bibinfo  {journal} {Journal of Statistical Mechanics: Theory and Experiment}\ }\textbf {\bibinfo {volume} {2007}},\ \bibinfo {pages} {P10004–P10004} (\bibinfo {year} {2007})}\BibitemShut {NoStop}%
\bibitem [{\citenamefont {Poilblanc}(2011)}]{Poilblanc2011}%
  \BibitemOpen
  \bibfield  {author} {\bibinfo {author} {\bibfnamefont {D.}~\bibnamefont {Poilblanc}},\ }\bibfield  {title} {\bibinfo {title} {Out-of-equilibrium correlated systems: Bipartite entanglement as a probe of thermalization},\ }\href {https://doi.org/10.1103/PhysRevB.84.045120} {\bibfield  {journal} {\bibinfo  {journal} {Phys. Rev. B}\ }\textbf {\bibinfo {volume} {84}},\ \bibinfo {pages} {045120} (\bibinfo {year} {2011})}\BibitemShut {NoStop}%
\bibitem [{\citenamefont {Fr\'erot}\ \emph {et~al.}(2018)\citenamefont {Fr\'erot}, \citenamefont {Naldesi},\ and\ \citenamefont {Roscilde}}]{Frerot2018}%
  \BibitemOpen
  \bibfield  {author} {\bibinfo {author} {\bibfnamefont {I.}~\bibnamefont {Fr\'erot}}, \bibinfo {author} {\bibfnamefont {P.}~\bibnamefont {Naldesi}},\ and\ \bibinfo {author} {\bibfnamefont {T.}~\bibnamefont {Roscilde}},\ }\bibfield  {title} {\bibinfo {title} {Multispeed prethermalization in quantum spin models with power-law decaying interactions},\ }\href {https://doi.org/10.1103/PhysRevLett.120.050401} {\bibfield  {journal} {\bibinfo  {journal} {Phys. Rev. Lett.}\ }\textbf {\bibinfo {volume} {120}},\ \bibinfo {pages} {050401} (\bibinfo {year} {2018})}\BibitemShut {NoStop}%
\bibitem [{\citenamefont {Li}\ and\ \citenamefont {Haldane}(2008)}]{paper:OG_ESpec}%
  \BibitemOpen
  \bibfield  {author} {\bibinfo {author} {\bibfnamefont {H.}~\bibnamefont {Li}}\ and\ \bibinfo {author} {\bibfnamefont {F.~D.~M.}\ \bibnamefont {Haldane}},\ }\bibfield  {title} {\bibinfo {title} {Entanglement spectrum as a generalization of entanglement entropy: Identification of topological order in non-abelian fractional quantum hall effect states},\ }\href {http://dx.doi.org/10.1103/PhysRevLett.101.010504} {\bibfield  {journal} {\bibinfo  {journal} {Phys. Rev. Lett.}\ }\textbf {\bibinfo {volume} {101}} (\bibinfo {year} {2008})}\BibitemShut {NoStop}%
\bibitem [{\citenamefont {Herviou}\ \emph {et~al.}(2016)\citenamefont {Herviou}, \citenamefont {Mora},\ and\ \citenamefont {Le~Hur}}]{paper:herviou2016}%
  \BibitemOpen
  \bibfield  {author} {\bibinfo {author} {\bibfnamefont {L.}~\bibnamefont {Herviou}}, \bibinfo {author} {\bibfnamefont {C.}~\bibnamefont {Mora}},\ and\ \bibinfo {author} {\bibfnamefont {K.}~\bibnamefont {Le~Hur}},\ }\bibfield  {title} {\bibinfo {title} {Phase diagram and entanglement of two interacting topological kitaev chains},\ }\href {http://dx.doi.org/10.1103/PhysRevB.93.165142} {\bibfield  {journal} {\bibinfo  {journal} {Phys. Rev. B}\ }\textbf {\bibinfo {volume} {93}} (\bibinfo {year} {2016})}\BibitemShut {NoStop}%
\bibitem [{\citenamefont {del Pozo}\ \emph {et~al.}(2023)\citenamefont {del Pozo}, \citenamefont {Herviou},\ and\ \citenamefont {Le~Hur}}]{paper:delpozo1}%
  \BibitemOpen
  \bibfield  {author} {\bibinfo {author} {\bibfnamefont {F.}~\bibnamefont {del Pozo}}, \bibinfo {author} {\bibfnamefont {L.}~\bibnamefont {Herviou}},\ and\ \bibinfo {author} {\bibfnamefont {K.}~\bibnamefont {Le~Hur}},\ }\bibfield  {title} {\bibinfo {title} {Fractional topology in interacting one-dimensional superconductors},\ }\href {https://doi.org/10.1103/PhysRevB.107.155134} {\bibfield  {journal} {\bibinfo  {journal} {Phys. Rev. B}\ }\textbf {\bibinfo {volume} {107}},\ \bibinfo {pages} {155134} (\bibinfo {year} {2023})}\BibitemShut {NoStop}%
\bibitem [{\citenamefont {del Pozo}\ \emph {et~al.}(2025)\citenamefont {del Pozo}, \citenamefont {Herviou}, \citenamefont {Dmytruk},\ and\ \citenamefont {Le~Hur}}]{paper:delpozo2}%
  \BibitemOpen
  \bibfield  {author} {\bibinfo {author} {\bibfnamefont {F.}~\bibnamefont {del Pozo}}, \bibinfo {author} {\bibfnamefont {L.}~\bibnamefont {Herviou}}, \bibinfo {author} {\bibfnamefont {O.}~\bibnamefont {Dmytruk}},\ and\ \bibinfo {author} {\bibfnamefont {K.}~\bibnamefont {Le~Hur}},\ }\bibfield  {title} {\bibinfo {title} {Model for topological $p$-wave superconducting wires with disorder and interactions},\ }\href {https://doi.org/10.1103/PhysRevB.111.075170} {\bibfield  {journal} {\bibinfo  {journal} {Phys. Rev. B}\ }\textbf {\bibinfo {volume} {111}},\ \bibinfo {pages} {075170} (\bibinfo {year} {2025})}\BibitemShut {NoStop}%
\bibitem [{\citenamefont {Metlitski}\ and\ \citenamefont {Grover}(2015)}]{Melitski2015}%
  \BibitemOpen
  \bibfield  {author} {\bibinfo {author} {\bibfnamefont {M.~A.}\ \bibnamefont {Metlitski}}\ and\ \bibinfo {author} {\bibfnamefont {T.}~\bibnamefont {Grover}},\ }\href {https://arxiv.org/abs/1112.5166} {\bibinfo {title} {Entanglement entropy of systems with spontaneously broken continuous symmetry}} (\bibinfo {year} {2015}),\ \Eprint {https://arxiv.org/abs/1112.5166} {arXiv:1112.5166} \BibitemShut {NoStop}%
\bibitem [{\citenamefont {Frérot}\ and\ \citenamefont {Roscilde}(2015)}]{Frerot2015}%
  \BibitemOpen
  \bibfield  {author} {\bibinfo {author} {\bibfnamefont {I.}~\bibnamefont {Frérot}}\ and\ \bibinfo {author} {\bibfnamefont {T.}~\bibnamefont {Roscilde}},\ }\bibfield  {title} {\bibinfo {title} {Area law and its violation: A microscopic inspection into the structure of entanglement and fluctuations},\ }\href {http://dx.doi.org/10.1103/PhysRevB.92.115129} {\bibfield  {journal} {\bibinfo  {journal} {Phys. Rev. B}\ }\textbf {\bibinfo {volume} {92}} (\bibinfo {year} {2015})}\BibitemShut {NoStop}%
\bibitem [{\citenamefont {Frérot}\ \emph {et~al.}(2017)\citenamefont {Frérot}, \citenamefont {Naldesi},\ and\ \citenamefont {Roscilde}}]{Frerot2017}%
  \BibitemOpen
  \bibfield  {author} {\bibinfo {author} {\bibfnamefont {I.}~\bibnamefont {Frérot}}, \bibinfo {author} {\bibfnamefont {P.}~\bibnamefont {Naldesi}},\ and\ \bibinfo {author} {\bibfnamefont {T.}~\bibnamefont {Roscilde}},\ }\bibfield  {title} {\bibinfo {title} {Entanglement and fluctuations in the {XXZ} model with power-law interactions},\ }\href {http://dx.doi.org/10.1103/PhysRevB.95.245111} {\bibfield  {journal} {\bibinfo  {journal} {Phys. Rev. B}\ }\textbf {\bibinfo {volume} {95}} (\bibinfo {year} {2017})}\BibitemShut {NoStop}%
\bibitem [{\citenamefont {Lemonik}\ and\ \citenamefont {Mitra}(2016)}]{Lemonik2016}%
  \BibitemOpen
  \bibfield  {author} {\bibinfo {author} {\bibfnamefont {Y.}~\bibnamefont {Lemonik}}\ and\ \bibinfo {author} {\bibfnamefont {A.}~\bibnamefont {Mitra}},\ }\bibfield  {title} {\bibinfo {title} {Entanglement properties of the critical quench of ${O}({N})$ bosons},\ }\href {https://doi.org/10.1103/PhysRevB.94.024306} {\bibfield  {journal} {\bibinfo  {journal} {Phys. Rev. B}\ }\textbf {\bibinfo {volume} {94}},\ \bibinfo {pages} {024306} (\bibinfo {year} {2016})}\BibitemShut {NoStop}%
\bibitem [{Note1()}]{Note1}%
  \BibitemOpen
  \bibinfo {note} {$\alpha $ is related to the so-called \protect \emph {slip-exponent} $\theta $ in previous works \cite {Maraga2015, Diehl2017}.}\BibitemShut {Stop}%
\bibitem [{\citenamefont {Peschel}\ and\ \citenamefont {Eisler}(2009)}]{Peschel2009}%
  \BibitemOpen
  \bibfield  {author} {\bibinfo {author} {\bibfnamefont {I.}~\bibnamefont {Peschel}}\ and\ \bibinfo {author} {\bibfnamefont {V.}~\bibnamefont {Eisler}},\ }\bibfield  {title} {\bibinfo {title} {Reduced density matrices and entanglement entropy in free lattice models},\ }\href {https://doi.org/10.1088/1751-8113/42/50/504003} {\bibfield  {journal} {\bibinfo  {journal} {Journal of Physics A: Mathematical and Theoretical}\ }\textbf {\bibinfo {volume} {42}},\ \bibinfo {pages} {504003} (\bibinfo {year} {2009})}\BibitemShut {NoStop}%
\bibitem [{\citenamefont {Botero}\ and\ \citenamefont {Reznik}(2004)}]{Botero2004}%
  \BibitemOpen
  \bibfield  {author} {\bibinfo {author} {\bibfnamefont {A.}~\bibnamefont {Botero}}\ and\ \bibinfo {author} {\bibfnamefont {B.}~\bibnamefont {Reznik}},\ }\bibfield  {title} {\bibinfo {title} {Spatial structures and localization of vacuum entanglement in the linear harmonic chain},\ }\href {https://doi.org/10.1103/PhysRevA.70.052329} {\bibfield  {journal} {\bibinfo  {journal} {Phys. Rev. A}\ }\textbf {\bibinfo {volume} {70}},\ \bibinfo {pages} {052329} (\bibinfo {year} {2004})}\BibitemShut {NoStop}%
\bibitem [{\citenamefont {Srednicki}(1993)}]{Srednicki1993}%
  \BibitemOpen
  \bibfield  {author} {\bibinfo {author} {\bibfnamefont {M.}~\bibnamefont {Srednicki}},\ }\bibfield  {title} {\bibinfo {title} {Entropy and area},\ }\href {https://doi.org/10.1103/PhysRevLett.71.666} {\bibfield  {journal} {\bibinfo  {journal} {Phys. Rev. Lett.}\ }\textbf {\bibinfo {volume} {71}},\ \bibinfo {pages} {666} (\bibinfo {year} {1993})}\BibitemShut {NoStop}%
\bibitem [{foo()}]{footnote1}%
  \BibitemOpen
  \href@noop {} {}\bibinfo {note} {The momentum cutoff $k_{\mathrm{max}} = 10\Lambda$ is taken sufficiently large compared to the regulator scale $\Lambda$ (see Eq.~(4)), so that the lattice dispersion $2 - 2\cos(ak)$ can be replaced by its continuum version $(ak)^2$, as used in the $\mathcal{O}(N)$ model, Eq.~(3).}\BibitemShut {Stop}%
\bibitem [{\citenamefont {Sharma}\ \emph {et~al.}(2022)\citenamefont {Sharma}, \citenamefont {J\"ager}, \citenamefont {Kraus}, \citenamefont {Roscilde},\ and\ \citenamefont {Morigi}}]{Roscilde2022}%
  \BibitemOpen
  \bibfield  {author} {\bibinfo {author} {\bibfnamefont {S.}~\bibnamefont {Sharma}}, \bibinfo {author} {\bibfnamefont {S.~B.}\ \bibnamefont {J\"ager}}, \bibinfo {author} {\bibfnamefont {R.}~\bibnamefont {Kraus}}, \bibinfo {author} {\bibfnamefont {T.}~\bibnamefont {Roscilde}},\ and\ \bibinfo {author} {\bibfnamefont {G.}~\bibnamefont {Morigi}},\ }\bibfield  {title} {\bibinfo {title} {Quantum critical behavior of entanglement in lattice bosons with cavity-mediated long-range interactions},\ }\href {https://doi.org/10.1103/PhysRevLett.129.143001} {\bibfield  {journal} {\bibinfo  {journal} {Phys. Rev. Lett.}\ }\textbf {\bibinfo {volume} {129}},\ \bibinfo {pages} {143001} (\bibinfo {year} {2022})}\BibitemShut {NoStop}%
\bibitem [{\citenamefont {Wang}\ \emph {et~al.}(2025)\citenamefont {Wang}, \citenamefont {Wang}, \citenamefont {Ding}, \citenamefont {Mao},\ and\ \citenamefont {Yan}}]{wang2025}%
  \BibitemOpen
  \bibfield  {author} {\bibinfo {author} {\bibfnamefont {Z.}~\bibnamefont {Wang}}, \bibinfo {author} {\bibfnamefont {Z.}~\bibnamefont {Wang}}, \bibinfo {author} {\bibfnamefont {Y.-M.}\ \bibnamefont {Ding}}, \bibinfo {author} {\bibfnamefont {B.-B.}\ \bibnamefont {Mao}},\ and\ \bibinfo {author} {\bibfnamefont {Z.}~\bibnamefont {Yan}},\ }\bibfield  {title} {\bibinfo {title} {Bipartite reweight-annealing algorithm of quantum monte carlo to extract large-scale data of entanglement entropy and its derivative},\ }\href {https://doi.org/10.1038/s41467-025-61084-7} {\bibfield  {journal} {\bibinfo  {journal} {Nature Communications}\ }\textbf {\bibinfo {volume} {16}},\ \bibinfo {pages} {5880} (\bibinfo {year} {2025})}\BibitemShut {NoStop}%
\bibitem [{Note2()}]{Note2}%
  \BibitemOpen
  \bibinfo {note} {It should be pointed out that this is by no means a unique choice to represent the evolution of the eigenvectors following the quench. This particular choice reflects the normalized eigenvectors of the $2N\times 2N$ symplectic correlation matrix $(iJC)$, however we have verified that the qualitative features presented in Fig.~\ref {fig:ballistic_vector} remain intact when plotting for example $|\phi _{q_{\parallel }}|^{2}$ or $|\Pi _{q_{\parallel }}|^{2}$ individually. Therefore, any combination of these two amplitudes will have the same qualitative features, making our choice a sufficient representation of the dynamical evolution in the modes.}\BibitemShut {Stop}%
\bibitem [{\citenamefont {Bornens}\ \emph {et~al.}(2026)\citenamefont {Bornens}, \citenamefont {Gliott},\ and\ \citenamefont {Cherroret}}]{Bornens2026}%
  \BibitemOpen
  \bibfield  {author} {\bibinfo {author} {\bibfnamefont {A.-S.}\ \bibnamefont {Bornens}}, \bibinfo {author} {\bibfnamefont {E.}~\bibnamefont {Gliott}},\ and\ \bibinfo {author} {\bibfnamefont {N.}~\bibnamefont {Cherroret}},\ }\href {https://arxiv.org/abs/2607.26620} {\bibinfo {title} {Critical non-thermal fixed point and the dynamical condensation phase transition}} (\bibinfo {year} {2026}),\ \Eprint {https://arxiv.org/abs/2607.26620} {arXiv:2607.26620 [cond-mat.quant-gas]} \BibitemShut {NoStop}%
\bibitem [{\citenamefont {Jian}\ \emph {et~al.}(2019)\citenamefont {Jian}, \citenamefont {Yin},\ and\ \citenamefont {Swingle}}]{Jian2019}%
  \BibitemOpen
  \bibfield  {author} {\bibinfo {author} {\bibfnamefont {S.-K.}\ \bibnamefont {Jian}}, \bibinfo {author} {\bibfnamefont {S.}~\bibnamefont {Yin}},\ and\ \bibinfo {author} {\bibfnamefont {B.}~\bibnamefont {Swingle}},\ }\bibfield  {title} {\bibinfo {title} {Universal prethermal dynamics in gross-neveu-yukawa criticality},\ }\href {https://doi.org/10.1103/PhysRevLett.123.170606} {\bibfield  {journal} {\bibinfo  {journal} {Phys. Rev. Lett.}\ }\textbf {\bibinfo {volume} {123}},\ \bibinfo {pages} {170606} (\bibinfo {year} {2019})}\BibitemShut {NoStop}%
\bibitem [{\citenamefont {Yin}\ and\ \citenamefont {Jian}(2021)}]{Yin2021}%
  \BibitemOpen
  \bibfield  {author} {\bibinfo {author} {\bibfnamefont {S.}~\bibnamefont {Yin}}\ and\ \bibinfo {author} {\bibfnamefont {S.-K.}\ \bibnamefont {Jian}},\ }\bibfield  {title} {\bibinfo {title} {Fermion-induced dynamical critical point},\ }\href {https://doi.org/10.1103/PhysRevB.103.125116} {\bibfield  {journal} {\bibinfo  {journal} {Phys. Rev. B}\ }\textbf {\bibinfo {volume} {103}},\ \bibinfo {pages} {125116} (\bibinfo {year} {2021})}\BibitemShut {NoStop}%
\bibitem [{Sci()}]{SciPy}%
  \BibitemOpen
  \bibfield  {title} {\bibinfo {title} {Initial value problem (ivp) solver in scipy \\ https://docs.scipy.org/doc/scipy/reference/generated\\/scipy.integrate.solve\_ ivp.html},\ }\href {https://docs.scipy.org/doc/scipy/reference/generated/scipy.integrate.solve_ivp.html} {\ }\BibitemShut {NoStop}%
\end{thebibliography}%

\appendix

\section{Numerical methods}
\label{app:num_methods}

Here, we provide additional details on the numerical methods used to solve the equations of motion and compute the entanglement spectrum. \newline

\noindent \textbf{Time integration and numerical solution of the equations of motion} 
To solve Eqs.~\eqref{eq:mode_func_eom} and \eqref{eq:effective_r}, we use Python and the initial-value solver provided by the``SciPy" \cite{SciPy} package. Specifically, we use the ``solve\textunderscore ivp'' function with the ``SecondOrderDerivative'' setting and the ``RK45'' method, with an absolute tolerance of $a_\text{tol}=10^{-12}$ and a relative tolerance of $r_\text{tol}=10^{-6}$. For the effective mass, we instead use Simpson's rule for numerical integration, also implemented in the SciPy package. 
We further performed a convergence test with respect to the numerical tolerances. Although the default tolerances in SciPy are several orders of magnitude smaller than those used in our calculations, we verified that $f_k(t)$ and $\dot{f}_k(t)$ are converged when using $a_\text{tol}=10^{-13}$ and $r_\text{tol}=10^{-7}$ for $r_i=100$, $u=10$, and $\delta=\pm1$ and $\delta=0$. The resulting momentum-space correlation functions, $\sim |f_k|^2$, $|\dot{f}_k|^2$, and $\sim f_k\dot{f}_k$, differ by at most $0.05\%$ in the IR and $0.37\%$ in the UV. This convergence test was performed for $\delta=-1$ at $t=200$, with $r_i=100$, $u=10$, and $\Lambda=\pi/2$. We also compared the spectra obtained for the quenches $\delta=0$ and $\delta=-1$, finding deviations of at most $0.06\%$. 
Finally, we use an equally spaced grid of $10^5$ radial momentum  $\smash{k=(k_x^2+k_y^2+k_z^2)^{1/2}}$ points.
We have verified that increasing the number of grid points does not affect the resulting mode functions.\newline

\noindent \textbf{Numerical implementation of the entanglement spectrum}
To obtain the entanglement spectrum, we compute the correlation matrices from the momentum-space mode functions $f_{k}$ and $\dot{f}_{k}$. Since the calculation is performed in a mixed momentum-space/real-space basis, the correlation functions $\langle \phi_{k}\phi_{-k}\rangle$, etc., are interpolated to obtain their values at the desired momenta, $\smash{(k_{z}^{2} + n^{2}\mathrm{d}q^{2})^{1/2}}$, where $k_z$ denotes the momentum in the direction along which we perform the Fourier transform, and $q_{\parallel}=n\mathrm{d}q$ is the discretized in-plane momentum. For the interpolation, we use the ``interp1d'' function from the SciPy package, with its default linear interpolation scheme. We then perform the Fourier transform using the ``fft'' function in Python.

To reduce the accumulation of numerical errors, we symmetrize the resulting correlation matrix $C_{q_{\parallel}}(i,j) = C_{q_{\parallel}}(|i - j|)$ explicitly. The correlation matrices are subsequently diagonalized using the ``numpy.linalg.eig'' function. The corresponding eigenvalues $\lambda_j$ are expected to be bounded from below by $1/2$. Small deviations from this bound are observed for $q_{\parallel}\gg 0$, where the ES approaches that of the initial state, $\lambda_j=1/2$, and for large $L$. For the results presented here, the bound is respected to within $10^{-5}$--$10^{-6}$, while the contribution of the zero mode remains well above this numerical uncertainty. We therefore conclude that our results for the IR properties of the ES are not affected by these numerical subtleties.

Finally, since our primary interest is in the IR properties, we use the quadratic, $\propto k^2$ dispersion for the mode functions. We have verified that enforcing periodicity in the Brillouin zone affects only the entanglement eigenvalues associated with the edges of the Brillouin zone, i.e., the UV properties. This further supports our conclusion that the IR properties of the entanglement spectrum are insensitive to the UV details of our implementation.

\section{Convergence and robustness of IR-limit results}
\label{app:numerics}

In this appendix, we describe the fitting procedures and demonstrate that the infrared results presented in Figs.~\eqref{fig:plateau_comp1},~\eqref{fig:plateau_comp2} and~\eqref{fig:entanglement_gap} are insensitive to the specific choice of the cutoff parameter $\Lambda\sim 1/a$ and to other microscopic details.\newline

\begin{figure}[h!]
\centering
\includegraphics[width= 0.87\linewidth]{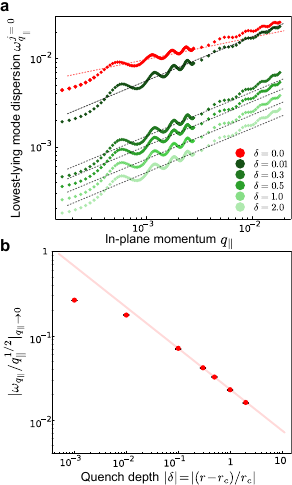}
\caption{
Panel a: Momentum dispersion of the lowest-lying entanglement mode ($j=0$) at low momentum, for different quench depths $\delta$, at a fixed long time $t=10^{3}$ and using a different set of cutoff parameters, $\Lambda=\pi/4$ and $k_{\mathrm{max}}=5\Lambda$, with $u=10$, $r_i=100$, and $N_s=60$.
Panel b: Prefactor of the entanglement dispersion as a function of the quench depth in the ordered phase. The solid red line corresponds to an algebraic fit of the form $a_1|\delta|^{b_1}$ for $|\delta|\geq 0.1$, with $a_1=0.023\pm3\times10^{-4}$ and $b_1=0.488\pm5\times10^{-3}$. The error bars on the data points in panel b correspond to the estimated uncertainties in the slopes extracted from panel a and are of order $10^{-3}$. The slopes in panel a were extracted by fitting the dispersion over momenta up to $\approx2\times10^{-3}$.
}
\label{fig:cutoff1}
\end{figure}

\noindent\textbf{IR properties of the entanglement spectrum}
Throughout the main text, we use the  cutoff parameters $\Lambda = \pi/2$ and $k_\text{max} = 10\Lambda$. Figure~\ref{fig:cutoff1}a shows that the infrared scaling $\smash{\sim q_{\parallel}^{\alpha}}$ of the lowest-lying entanglement mode dispersion is unaffected by an alternative choice, namely $\Lambda = \pi/4$ and $k_\text{max} = 5\Lambda$. Only the overall magnitude of the entanglement mode depends on the UV cutoff.
Similarly, Fig.~\ref{fig:cutoff1}b shows that the prefactor of the entanglement dispersion, obtained for a different choice of the cutoff parameters, exhibits the same critical scaling with $|\delta|=|(r-r_c)/r_c|$ as that shown in Fig.~\ref{fig:plateau_comp2}. The error bars on the individual data points are again negligible and are therefore not visible in the plot.
It should be noted that the characteristic  time beyond which the entanglement dispersion approaches its asymptotic value depends on the UV cutoff. For $\delta\to0$, the relaxation time becomes too long to reach the asymptotic regime within our computational capabilities, both in the main text and in the present analysis. We can therefore only infer that, at sufficiently long times, the data points converge towards the same algebraic curve.\newline

\noindent\textbf{Fitting procedures:}
The estimated errors on the IR slopes of $\smash{\omega_{q_\parallel}/q_\parallel^{1/2}}$ are several orders of magnitude smaller than the corresponding fitted values. Neglecting these errors yields negligible $\chi^2_{\mathrm{red}}$ and covariance, reflecting the excellent agreement of the data with the algebraic scaling, see Fig.~\ref{fig:plateau_comp2} and panel b of Fig.~\ref{fig:cutoff1}. 
Including the errors does not significantly change the fitted values, but makes their estimated covariance highly sensitive to both the momentum range used to extract the slopes and the $\delta$ range used for the algebraic fit. The resulting $\chi^2_{\mathrm{red}}$ can also vary substantially, reaching values as large as $\chi^2_{\mathrm{red}}\approx40$ in Fig.~5. These sensitivities prevent a reliable quantitative interpretation of the fitting uncertainties.\newline

\begin{figure}[h!]
\centering
\includegraphics[width=0.9\linewidth]{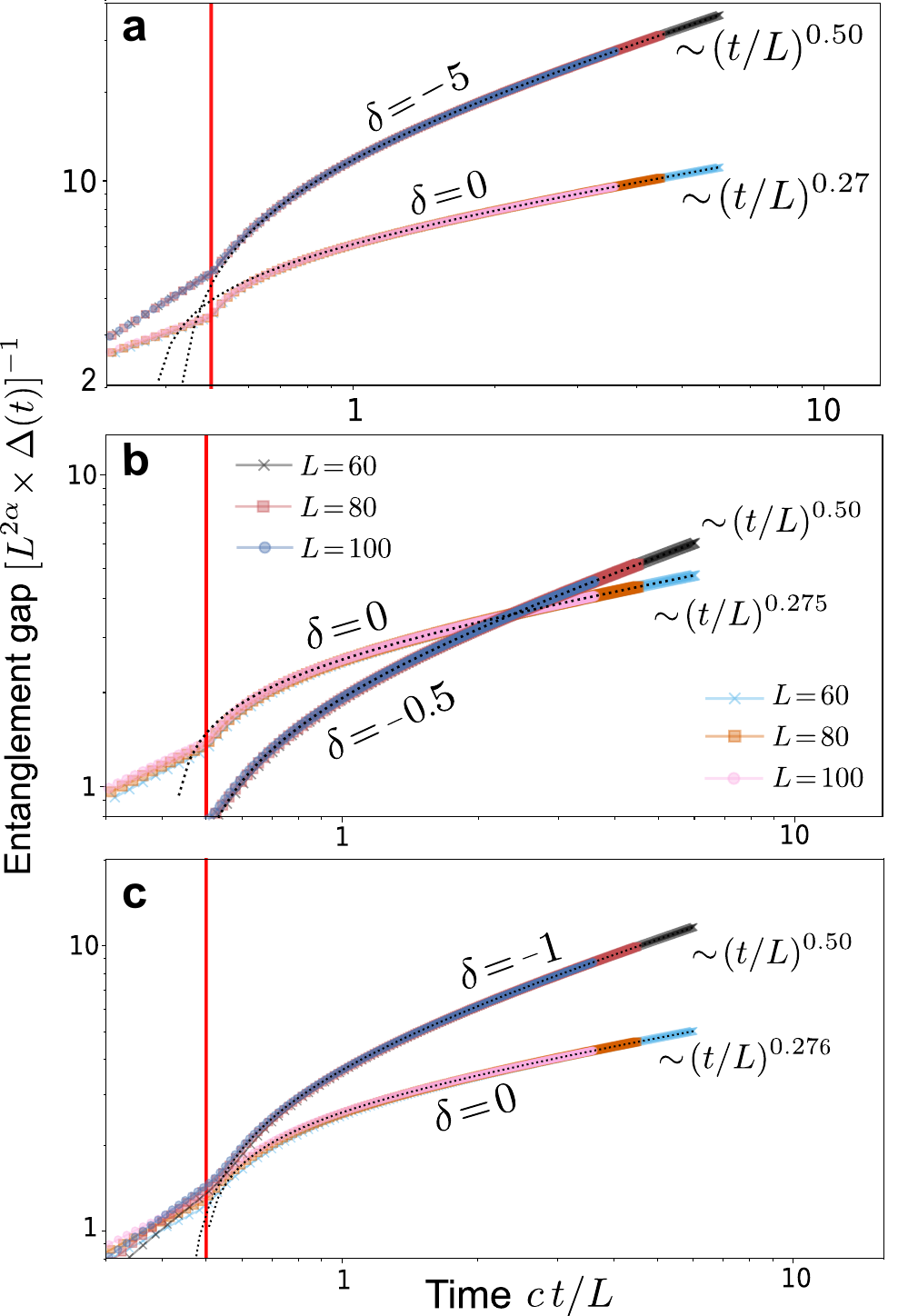}
\caption{
Inverse of the entanglement gap $\smash{\Delta(t)\equiv\omega_{q_\parallel=0}^{j=0}(t)}$ as a function of time for three subsystem sizes $L$, shown for quenches at ($\delta=0$) and below ($\delta<0$) the critical point. The three panels use the cutoff parameters $\Lambda = \pi/4$, $k_{\mathrm{max}} = 5\Lambda$, different from those in Fig.~\ref{fig:entanglement_gap}. Panel a: $u=10$ and $r_i=100$ , Panel b: $u=10$ and $r_i=10$; Panel c: $u=5$ and $r_i=10$. All curves obey the scaling law \eqref{eq:gapscaling} of the main text, as well as the asymptotic relation \eqref{eq:Wscaling} beyond the light-cone boundary $t = L/(2c)$ (vertical line), demonstrating that these laws are robust against parameter changes. 
In panels a, b and c, the vertical axis is rescaled by $L^{2\alpha}$, respectively with $\alpha=0.267, 0.288, 0.312$ for the case $\delta =0$, and $\alpha=0.50, 0.510, 0.513$ for the case $\delta<0$. We use a total number of point $N_{\mathrm{tot}} = 5\times 10^{5}$.
}
\label{fig:gap_variedparam}
\end{figure}

\noindent\textbf{Entanglement gap scaling for different parameters}
We now turn to the entanglement gap, i.e., Fig.~\eqref{fig:entanglement_gap} of the main text,  which we have further investigated for different values of the microscopic parameters. Our results are presented in Fig.~\ref{fig:gap_variedparam}, where we show the entanglement gap for a different choice of the cutoff parameters $(\Lambda,k_\text{max})$ (panel a), a different initial mass $r_i$ (panel b), and a different interaction strength $u$ (panel c).
These three panels demonstrate that both the scaling law  \eqref{eq:gapscaling} of the main text and the long-time asymptotic relation \eqref{eq:Wscaling} for the scaling function are robust with respect to variations in the microscopic parameters, as expected. In all three cases, an excellent collapse of the data is obtained for values of the scaling exponent $\alpha$ very close to the value expected for quenches to or below the critical point.

\begin{figure}[h!]
\centering
\includegraphics[width =0.82\linewidth]{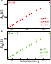}
\caption{Entanglement entropy $S_{q_{\parallel}}(L)$ as a function of the subsystem size $L$ for individual $q_{\parallel}$ modes,  $q_{\parallel}= 0, 0.25, 0.5$, following a quench a) to  ($\delta = 0$) and b) below ($\delta=-1$) the critical point. Both panels are shown at fixed time $t = 1000$. Numerical parameters are the same as in Fig.~\ref{fig:entropy_growth_log}.}
\label{fig:dom1}
\end{figure}

\section{Long-time scaling of the EE}
\label{app:EE_scaling}

In this appendix, we present a detailed  numerical scaling analysis that justifies Eq.~\eqref{eq:entropy_scaling} of the main text:
\begin{equation}
\label{eq:fullEE_scaling}
    S(L|L_\text{tot}-L, t) \simeq AL_{\parallel}^{2}L + BL_{\parallel}^{2} + \alpha\log(t) + \dots
\end{equation}
which is valid at long times, $ct\gg L$. This scaling form follows from the linear growth of the entanglement entropy in each individual $q_{\parallel}$ sector [see Eq.~\eqref{eq:full_EE}] as
\begin{equation}
\label{eq:scaling_qblock}
    S_{q_{\parallel}}(L) = a_{q_{\parallel}}L +\delta_{q_\parallel,0}\,\alpha\ln t +b_{q_\parallel}\ldots
\end{equation}
where the coefficients $a_{q_{\parallel}}$ and $b_{q_{\parallel}}$ are found to depend only weakly on $q_{\parallel}$. Summing over all $q_{\parallel}$ sectors then naturally leads to Eq.~\eqref{eq:fullEE_scaling}.\newline

To begin with, we investigate the first, leading-order term on the right-hand side of Eq.~\eqref{eq:scaling_qblock}. This term, which gives rise to the dominant volume-law contribution in Eq.~\eqref{eq:fullEE_scaling}, is linear in $L$. 
\begin{figure}[h!]
\centering
\includegraphics[width =0.88\linewidth]{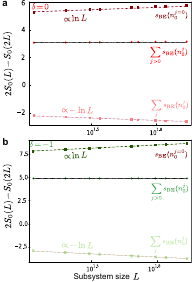}
\caption{Contribution of the $q_{\parallel} = 0$ sector, $S_{q_{\parallel} = 0}(L)$, to the EE as a function of system size, following a quench a) to  ($\delta = 0$) and b) below ($\delta=-1$) the critical point. 
Both panels are shown at fixed time $t = 1000$. 
The plots display the difference $2S_0(L) - S_0(2L)$, which eliminates the $\propto L$ term in Eq.~\eqref{eq:scaling_qblock}. In each panel, dark (light) squares show the contribution from the $j=0$ ($j>0$) modes only. These contributions scale as $(-)\ln L$, as indicated by the dashed lines. As a result, the contribution from all $j$ modes (middle dots) is nearly independent of $L$. Numerical parameters are the same as in Fig.~\ref{fig:entropy_growth_log}.}
\label{fig:subdom1}
\end{figure}
This scaling is demonstrated in Fig.~\ref{fig:dom1} for both quenches to (panel a) and below (panel b) the critical point. Note that the linear scaling is observed for both $q_{\parallel}=0$ and $q_{\parallel}\ne 0$, with very little variation of the slope $a_{q_{\parallel}}$ as a function of $q_{\parallel}$. This weak dependence on $q_{\parallel}$ justifies the form of Eq.~\eqref{eq:fullEE_scaling} upon summing over all $q_{\parallel}$ sectors.\newline

Next, we examine the contribution to $S_{q_\parallel}(L)$ that is sublinear in $L$, again at fixed time. To this end, we consider the difference $2S_{q_\parallel}(L,t) - S_{q_\parallel}(2L,t)$ which, by construction, eliminates the linear contribution in Eq.~\eqref{eq:scaling_qblock}, while retaining any potential other term that would scale polynomially or logarithmically with $L$. We first consider the $q_{\parallel}=0$ sector in Fig.~\ref{fig:subdom1}, examining separately the contributions from the $j=0$ and $j>0$ modes [recall that $S_{q_\parallel}=\sum_j s_\text{BE}(n_{q_\parallel}^j)$]. As found in the main text, see Eq.~\eqref{eq:log_s0}, we recover the $\alpha\ln L$ scaling of the $j=0$ contribution. However, Fig.~\ref{fig:subdom1} also reveals that the contribution from all $j>0$ modes exhibits the opposite behavior, scaling as $-\alpha\ln L$. Consequently, after summing over all $j$ modes, the entropy $S_{q_\parallel=0}(L)$ is found to be nearly independent of $L$ (more precisely, for $\delta = 0$, we find a residual contribution $S_0(L)\sim 0.04\log L$, which we nevertheless attribute to finite-time and finite-size effects).
\begin{figure}[h!]
\centering
\includegraphics[width =0.87\linewidth]{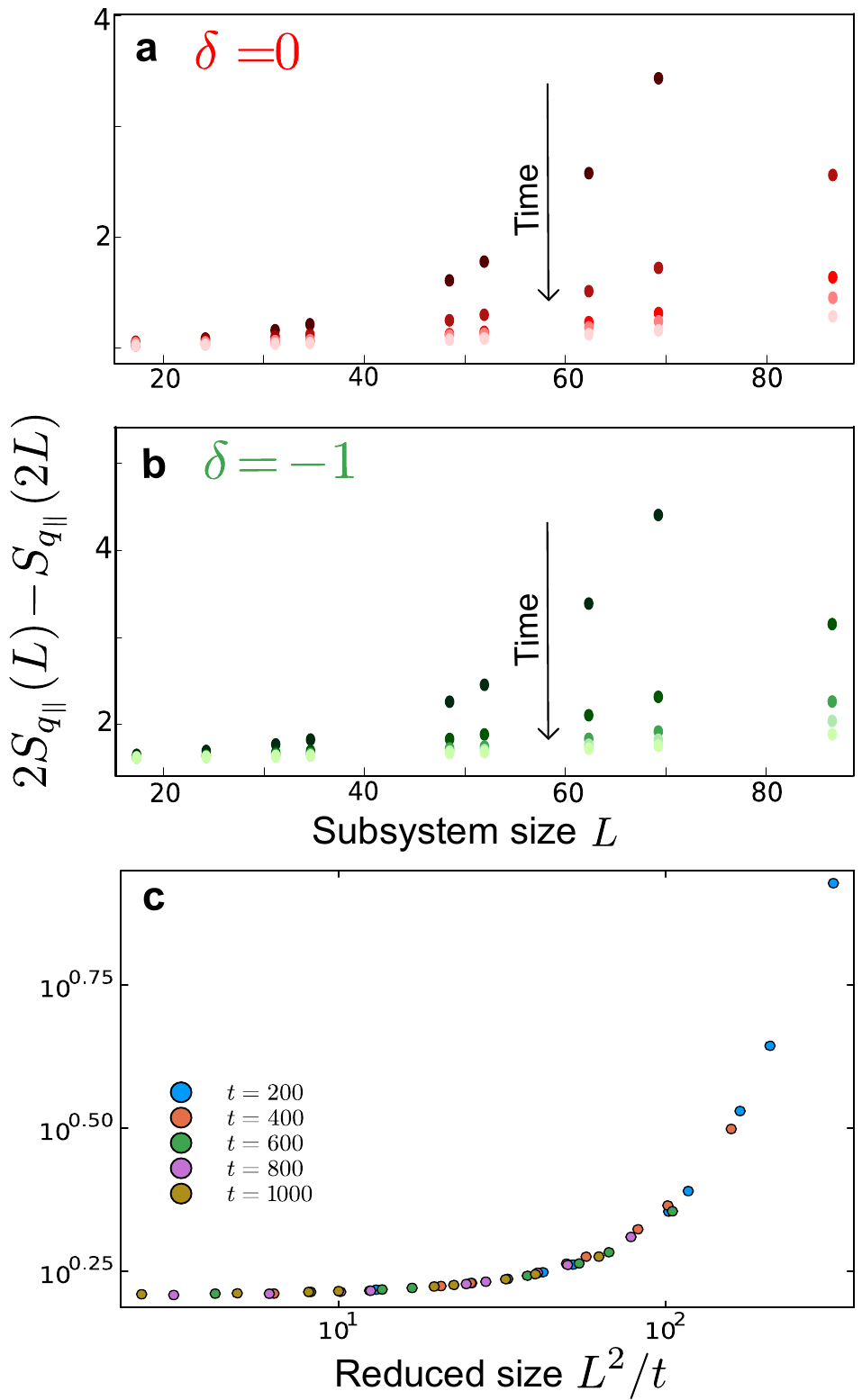}
\caption{Difference $2S_{q_\parallel}(L)-S_{q_\parallel}(2L)$ as a function of $L$, shown here for $q_{\parallel}\approx0.25$, following a quench a) to ($\delta=0$) and b) below ($\delta=-1$) the critical point. Dots with different shades correspond to different times, $t\approx200,400,600,800,$ and $1000$, from top to bottom. Panel c shows a rescaling of the data for $\delta=-1$, demonstrating that the finite-time corrections to $S_{q_\parallel}$ scale as $L^2/t$, thus becoming independent of $L$ when extrapolated to long times. Numerical parameters are the same as in Fig.~\ref{fig:entropy_growth_log}.}
\label{fig:subdom2}
\end{figure}

The case $q_\parallel\ne 0$ is addressed in Fig.~\ref{fig:subdom2}a-b, where we observe a more pronounced dependence of $2S_{q_\parallel}(L))-S_{q_\parallel}(2L)$ on $L$ than for $q_{\parallel}=0$ at finite time. These finite-time corrections nevertheless seem to disappear in simulations performed at longer times. In Fig.~\ref{fig:subdom2}c we show the same data as in Fig.~\ref{fig:subdom2}b,  now plotted as a function of $L^2/t$. The resulting data collapse demonstrates that the finite-time corrections to $S_{q_\parallel}$ become independent of $L$ when extrapolated to the long-time limit. We also verified that the same rescaling collapses the data for $\delta =0$ in Fig.~\ref{fig:subdom2}a). We therefore conjecture that those do not persist in the long-time expansion of the EE, Eq.~\eqref{eq:scaling_qblock}. 
\newline
\begin{figure}[H]
\centering
\includegraphics[width =1\linewidth]{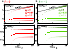}
\caption{Entanglement entropy $S_{q_\parallel}=\sum_j s_\text{BE}(n_{q_\parallel}^j)$ as a function of time at fixed $L=48$, following a quench a) to ($\delta=0$) and b) below ($\delta=-1$) the critical point. The upper panels show the contribution from the $j=0$ mode, while the lower panels show the combined contribution from all $j>0$ modes. The results show that the $\alpha\ln t$ term in Eq.~\eqref{eq:scaling_qblock} originates exclusively from the $(q_\parallel=0,j=0)$ mode. Numerical parameters are the same as in Fig.~\ref{fig:entropy_growth_log}.}
\label{fig:dom2}
\end{figure} 
Finally, we show in Fig.~\ref{fig:dom2} that the subleading, $L$-independent contribution in Eq.~\eqref{eq:scaling_qblock} indeed takes the form $\delta_{q_\parallel,0}\alpha\ln t+b_{q_\parallel}$. To disentangle the contributions of the different $j$ modes to $S_{q_\parallel}=\sum_j s_\text{BE}(n_{q_\parallel}^j)$, we examine the time dependence of the $j=0$ and $j>0$ contributions separately. For $q_\parallel\ne0$, we find no evidence of time growth for any $j$, either for quenches to (panel a) or below (panel b) the critical point. In contrast, for $q_\parallel=0$, we find that the $j=0$ contribution grows as $\alpha\ln t$, while the $j>0$ contributions remain time independent, as already shown in the main text. This confirms the scaling form stated above.

\end{document}